\documentclass[aps,preprintnumbers,superscriptaddress,twocolumn,amsmath,amssymb,floatfix,prb]{revtex4-1} 
\pdfoutput=1

\usepackage{graphicx,float}
\usepackage[usenames,dvipsnames]{color}
\usepackage[colorlinks,bookmarks=false,citecolor=NavyBlue,linkcolor=Red,urlcolor=blue]{hyperref}
\usepackage{enumitem} 

\usepackage[export]{adjustbox}

\usepackage{mathtools}

\newcommand\be{\begin{equation}}
\newcommand\bea{\begin{eqnarray}}
\newcommand\bes{\begin{subequations}}
\newcommand\esu{\end{subequations}}
\newcommand\ee{\end{equation}}
\newcommand\eea{\end{eqnarray}}

\newcommand{\cmmnt}[1]{}

\newcommand\ba         {\begin{eqnarray} } 
\newcommand\ea         {\end{eqnarray} } 

\usepackage{braket}
\usepackage[normalem]{ulem}

\newcommand\ocite[1]{[\onlinecite{#1}]}
\newcommand{\dd}{{\rm d}}

\usepackage{dsfont}

\definecolor{teal}{rgb}{0.0, 0.5, 0.5}

\begin{document}

\title{Quantum corrections to the classical field approximation for one-dimensional quantum many-body systems in equilibrium}

\author{Alvise Bastianello}
\affiliation{Institute for Theoretical Physics, University of Amsterdam, Science Park 904, 1098 XH Amsterdam, The Netherlands}
\author{Maksims Arzamasovs}
\affiliation{Department of Applied Physics, School of Science, Xi'an Jiaotong University, Xi'an 710049, Shaanxi, China}
\affiliation{Shaanxi Province Key Laboratory of Quantum Information and Quantum Optoelectronic Devices,
Xi'an Jiaotong University, Xi'an 710049, Shaanxi, China}
\author{Dimitri M. Gangardt}
\affiliation{School of Physics and Astronomy, University of Birmingham, Edgbaston, Birmingham, B15 2TT, United Kingdom}

\date{\today}

\pacs{}

\begin{abstract}
  We present a semiclassical treatment of one-dimensional many-body quantum
  systems in equilibrium, where quantum corrections to the classical field
  approximation are systematically included by a renormalization of the
  classical field parameters. Our semiclassical approximation is reliable in
  the limit of weak interactions and high temperatures.  As a specific
  example, we apply our method to the interacting Bose gas and study
  experimentally observable quantities, such as correlation functions of
  bosonic fields and the
  full counting statistics of the number of particles in an interval. Where
  possible, our method  is checked against exact results
  derived from integrability, showing excellent agreement.
\end{abstract}

\maketitle

\section{Introduction}
\label{sec_intro}

As experimental 
techniques in the field of ultra cold atoms reach their maturity
\ocite{RevModPhys.80.885,Cazalilla_2010,RevModPhys.83.1405,RevModPhys.83.863},
the characterization of these quantum many-body systems in terms of their
correlation properties becomes an important and timely issue. Indeed, in most
of the experiments the information about the state of the system and its
macroscopic parameters are inferred from one or two-particle correlation
functions: the momentum distribution and higher momentum
correlations \ocite{Richard2003,PhysRevA.86.043626,Fang2016}, phase
correlations \ocite{Langen2013}, local density correlations
\ocite{Tolra_04,PhysRevLett.95.190406,Haller2011}, or the density-density correlation functions \ocite{PhysRevA.70.013603}.  Recent
experiments \ocite{Schweigler2017,schweigler2020decay} aim to measure higher-order correlation
functions to provide valuable insight into effects of the interactions and
require theoretical predictions beyond perturbation theory.

Quite expectedly, the studies of many-body correlations has been especially
numerous in one-dimensional interacting models where, due to kinematic
constraints, interaction effects are always strong and standard perturbative
techniques are thus of limited applicability.  On the other hand, there are
several powerful nonperturbative techniques available in 1D.  For low
temperatures, the method of bosonization
\ocite{haldane1981demonstration,haldane1981luttinger,haldane1981effective,RevModPhys.83.1405}
using collective hydrodynamic description of the many particle degrees of
freedom  proved to be extremely valuable in characterizing long-distance, long-time
asymptotics of correlation functions. However, bosonization breaks down at higher temperatures and one should revert to other
approaches.  

Fortunately, certain one-dimensional models offer exact
solutions due to their integrability \ocite{MattisBook,korepin1997quantum}.
In these  systems, some experimentally
relevant observables can be exactly computed for arbitrary temperatures and
interactions, both in \ocite{takahashi2005thermodynamics} and out of
equilibrium \ocite{Calabrese_2016}.  An outstanding example is the
Lieb-Liniger (LL) model \ocite{PhysRev.130.1605,PhysRev.130.1616}, which
describes bosons with contact pairwise interactions. This model has proven to
be an excellent description of experiments with one-dimensional bosons
\ocite{PhysRevLett.81.938,Kinoshita1125,PhysRevLett.95.190406,Kinoshita2006,PhysRevLett.100.090402,PhysRevA.83.031604,PhysRevLett.115.085301,PhysRevA.91.043617,PhysRevLett.122.090601}
and will be the example of an interacting many-body system
discussed in this publication.

However, despite its appeal, integrability has its limitations:
numerous relevant systems are not exactly solvable and, even when they are,
several experimentally relevant quantities are out of reach of the
state-of-the-art integrable techniques. For instance, two-point correlation
functions are only partially controlled \ocite{CortesCubero2019,10.21468/SciPostPhys.8.1.004,De_Nardis_2015,Pozsgay2018,Seel_2007,Kozlowski_2011}
and their exact calculation is still an open problem.

Analytic approaches aside, 1D interacting systems in the continuum
present a challenge for numerical treatments as well. In contrast with lattice
models,  the continuum limit is notoriously hard to be accessed with the
density matrix renormalization group (DMRG) methods \ocite{Schollwock2013}.
When applicable, integrability-based numerical methods
 exist \ocite{doi:10.1063/1.3216474,Caux_2007} and describe the low energy
states reliably, but their efficiency is drastically reduced at higher
temperatures.

On the other hand, it is precisely in this regime that quantum systems are
amenable to semiclassical methods
\ocite{PhysRevA.96.013623,PhysRevA.90.033611,PhysRevA.86.043626,Adhikari_2002,PhysRevA.98.023613,De_Nicola_2019,MUSSARDO2007101,Mussardo_2015}:
in the limit of high temperatures and weak interactions, the modes of the
system are macroscopically occupied and quantum fluctuations can be neglected
in favor of thermal ones. In the context of ultra-cold
atoms this approach, known as \emph{the classical field method}, was pioneered 
in  studies of equilibrium and nonequilibrium physics of Bose-Einstein
condensates in higher dimensions \ocite{Brewczyk_2007,Blakie2008,Cockburn2009}.
It leads to stochastic Langevin-type equations for bosonic fields, to
be simulated numerically. 

For one-dimensional bosons the classical field
approach can be complemented by the transfer
matrix method \ocite{PhysRevB.6.3409} which allows for the computation
of one- and two-point correlation functions
\ocite{doi:10.1080/09500340008232189} and, more recently, calculation of the momentum
correlations \ocite{PhysRevA.86.033626} in the LL model.  Very recently,
combining the classical field and transfer matrix approaches allowed two of us
to access the entire probability distribution of the particle number on an
interval of arbitrary length, known as the full counting statistics (FCS)
\ocite{PhysRevLett.122.120401}. 
The FCS contains  much more
information than the expectation values of the moments of an operator,
and it attracted a lot of interest in various contexts
\ocite{Cherng_2007,Bortz_2007,PhysRevLett.100.165706,PhysRevE.87.022114,Shi_2013,PhysRevLett.111.080402,PhysRevB.95.035119,
  Collura_2017,PhysRevB.96.235109,najafi2019formation,PhysRevLett.119.236401,Gritsev2006,PhysRevLett.110.060602,10.21468/SciPostPhys.4.6.043,PhysRevB.101.041110,PhysRevB.101.041110,10.21468/SciPostPhys.7.6.072,PhysRevB.88.134301,
  PhysRevLett.120.190601,Bastianello_2018,10.21468/SciPostPhys.8.3.036,vecchio2020exact}. Moreover, semiclassical approximations have been successfully merged together with integrability in several instances to study in- and out- of equilibrium protocols \ocite{De_Luca_2016,PhysRevLett.123.130602,10.21468/SciPostPhys.4.6.045,vecchio2020exact}.

However, at any finite temperature quantum fluctuations on top of the
classical result still do matter. So far, the systematic treatment of such
quantum corrections has been absent and the present work is intended to fill
this gap.  We show how deviations from the classical limit can be
systematically accounted for through a proper renormalization of the parameters of
the classical energy functional and observables. Then, the observables of such
effective classical model can be reliably numerically computed either with the transfer
matrix approach
\ocite{PhysRevB.6.3409,Krumhansl1975,doi:10.1080/09500340008232189} or the
Metropolis-Hastings algorithm
\ocite{10.1093/biomet/57.1.97,doi:10.1080/00031305.1995.10476177}.  This
approach offers an easy interpretation of the effect of quantum fluctuations
and their relative importance in different temperature regimes relevant to
experiments. We emphasize that our approach is essentially different from the
standard mapping of 1D quantum systems onto two-dimensional classical systems since directly tackling a 2D problem with Monte Carlo
methods, or otherwise, is challenging, while the effective 1D classical system
is accessed relatively straightforwardly.
As we will clarify later on, the small parameter of the expansion is not the inverse temperature, despite the expansion being valid in the high-temperature limit. 
Rather, we expand in the relative strength of
quantum fluctuations compared to the classical ones.
The effective classical action is extracted from the quantum problem through the following steps: \emph{i)} spotting and isolating the classical degree of freedom from the quantum fluctuations, \emph{ii)} integrating out the quantum variables in a perturbative manner, and \emph{iii)} properly reorganizing the perturbative series in an effective action for the classical degree of freedom. The reader can refer to Ref. \ocite{kleinert2009path} for the single-particle physics, while in this work we generalize this approach to the many-body case.

The paper is organized as follows.  In order to provide a short pedagogical
introduction to our method, in Sec. \ref{sec_singleboson}, we present a simple
toy model consisting of a single degree of freedom, namely the anharmonic
oscillator. Quantum corrections  on top of the
classical result are considered and   compared with the first principle exact solution to the quantum mechanical problem, showing an
excellent agreement.

In Sec. \ref{sec_Bosegas}, we move to proper many-body systems and, for the
sake of concreteness, focus on the Lieb-Liniger model. Before embarking on the
semiclassical expansion for this model we provide a short summary of its
integrability: exact results available for the LL model can serve as
benchmarks of the semiclassical expansion. It should be noted, however, that the
semiclassical expansion does not rely on integrability and hence is of wide
applicability. The one-particle density matrix and the FCS of the number of
particles, for which no exact results are available, are studied within our
semiclassical approach. 
In Sec. \ref{sec_conclusions}, we gather our
conclusions, outline possible future research directions and discuss the relevance of our results for experiments. Details of the
numerical methods used to deal with the 1D effective classical model are given
in two appendices.

\section{Semiclassical approach to the anharmonic oscillator}
\label{sec_singleboson}

Before turning to many-body systems Ð the main focus of this paper Ð it is
useful to introduce the semiclassical expansion formalism using  a simple toy
model, namely, the anharmonic oscillator \ocite{kleinert2009path}.
This will allow to keep technical details at the bare minimum
later on. 
Let us consider an anharmonic quantum oscillator in thermal equilibrium
at temperature $T$, described by the density matrix
$\hat{\rho}\propto e^{-\beta \hat{H}}$, with $\beta=1/T$.  The oscillator is
governed by the following quantum Hamiltonian \be\label{eq_anhq}
\hat{H}=\frac{\hat{p}^2}{2}+\frac{\hat{q}^2}{2}+\frac{c}{4!}\hat{q}^4\, , \ee
where $\hat{p}$ and $\hat{q}$ are canonically conjugate Hermitian operators
$[\hat{q},\hat{p}]=i$. Hereafter, we use $\hbar=1$, $k_B=1$ unless stated
otherwise.

The classical limit of Hamiltonian \eqref{eq_anhq} is expected to emerge from
quantum mechanics in the limit of large occupation numbers, which is achieved
at high temperatures.  However, since the typical value of the coordinate scales as
$\hat{q}\sim \sqrt{T}$, a naive increase of temperature enhances the role of
the interactions in Eq. \eqref{eq_anhq} as well, making the problem
intrinsically quantum.  Keeping the nonlinear term of the same order as the
linear ones in the classical limit $\beta\to 0$, one should also require
$c \to 0$. Thus the classical limit is achieved in the high-temperature/weak
interaction limit.
More precisely, consider  the classical anharmonic oscillator 
\be\label{eq_anhcl}
H=\frac{p^2}{2}+\frac{q^2}{2}+ \frac{c_\text{cl}}{4!} q^4\, ,
\ee
with  $p$ and $q$ being classical
conjugated variables (notice the absence of the operator ``hats")
with the  ``classical interaction strength'' $c_\text{cl}=c\beta^{-1} = c T$.  

Now
let $\mathcal{O}(x)$ be an analytic function of its argument, then let us
consider quantum observables in the form $\mathcal{O}[\sqrt{\beta} \hat{q}]$,
where the $\beta-$dependence is inserted to achieve a well-defined
semiclassical limit.  Then, in the high-temperature/weak coupling limit one
has the following correspondence: \be\label{eq_cllimit} \lim_{\beta \to 0
}\frac{1}{\mathcal{Z}_{\text{q}}}\text{Tr}[\mathcal{O}[\sqrt{\beta}\hat{q}]
e^{-\beta \hat{H}}]=\frac{1}{\mathcal{Z}}\int \dd q\, \mathcal{O}[q]
e^{-\frac{q^2}{2}-\frac{c_\text{cl}}{4!} q^4}\, , \ee where
$\mathcal{Z}_{\text{q}}$ and $\mathcal{Z}$ are the classical and quantum partition
functions, respectively. On the right we recognize the expectation value in a
classical thermal ensemble with classical energy Eq. \eqref{eq_anhcl} (with
the momentum contribution having been integrated out, since the focus is on
$q-$ dependent observables).

The right-hand side of Eq. \eqref{eq_cllimit} becomes accurate only in the
classical field limit: at any finite temperature quantum corrections
will affect the expectation values of the operators. We
are now going to show how this can be captured by means of a suitable
renormalization of the classical energy and observables \be\label{eq_rencl}
\frac{1}{\mathcal{Z}_{\text{q}}}\text{Tr}[\mathcal{O}[\sqrt{\beta}\hat{q}]
e^{-\beta \hat{H}}]=\frac{1}{\mathcal{Z}}\int \dd q\,
\mathcal{O}_\text{eff}[q] e^{-\mathcal{S}_\text{eff}[q]}\, .  \ee The
effective observable $\mathcal{O}_\text{eff}$ and action
$\mathcal{S}_\text{eff}$ are explicitly temperature-dependent: hereafter we
provide the systematic expansion of these effective quantities around the
classical limit.

As the first step, we need to express the quantum expectation value in the form Eq. \eqref{eq_rencl}: in order to do so, we consider the path integral formulation of thermal averages. We introduce a real field $q_E(\tau)$ with $\tau \in[0,\beta]$ being the Euclidean time: the left-hand side (l.h.s.) of Eq. \eqref{eq_rencl} can be exactly rewritten as
\be\label{eq_path_quantum}
\int \mathcal{D}q_E\, \mathcal{O}[\sqrt{\beta}q_E(0)]e^{-\int_0^\beta \dd \tau\left\{ \, \frac{(\partial_\tau q_E)^2}{2}+\frac{q_E^2}{2}+\frac{c}{4!}q_E^4\right\}}\, ,
\ee
where we neglect an unimportant overall normalization constant and periodic boundary conditions are enforced on the Matsubara interval $\tau\in [0,\beta]$.
Intuitively, in the high-temperature limit the Matsubara interval shrinks to a point, hence the oscillations of the field $q_E(\tau)$ are expected to be suppressed. Indeed, if the $\tau-$derivative is neglected in the expression above, the classical energy functional is recovered.
This statement is made more rigorous by expanding the field $q_E$ in terms of its Fourier components, which we refer to as Matsubara modes
\be\label{eq_E_mat}
q_E(\tau)=\frac{q}{\sqrt{\beta}}+\sum_{n\ne 0} e^{i2\pi n \tau/\beta}\frac{q_n}{\sqrt{\beta}}\, .
\ee
As the notation suggests, the $n=0$ Matsubara mode can be interpreted as the emergent classical degree of freedom, while the modes $q_{n\ne 0}$ are responsible for the quantum fluctuations. Notice that since the Euclidean field is real by definition we have $q_n=q_{-n}^*$.
We can now express the path integral Eq. \eqref{eq_path_quantum} in terms of the Matsubara modes, isolating the contribution of the classical field from the rest,
\be
\int \dd q\, e^{-\mathcal{S}_\text{cl}[q]}\int \dd q_{n\ne 0}\, \mathcal{O}\left[\sum_n q_n\right]e^{-\mathcal{S}_\text{free}[q_{n\ne 0}]-\mathcal{S}_\text{int}[q,q_{n\ne 0}]}\, .
\ee
The Euclidean action above is split into three parts: the classical part $\mathcal{S}_\text{cl}$ which depends only on the classical degree of freedom, the free action $\mathcal{S}_\text{free}$ for the nonzero Matsubara modes, and the interacting part.
The three terms explicitly read
\begin{eqnarray}
&&\mathcal{S}_\text{cl}[q]=\frac{q^2}{2}+\frac{c_\text{cl}}{4!}q^4\, ,\\
&&\mathcal{S}_\text{free}[q_{n\ne 0}]= \sum_{n>0} [(2\pi n \beta^{-1})^2+1]|q_n|^2\, ,\\
&&\nonumber\mathcal{S}_\text{int}[q,q_{n\ne 0}]= \frac{c_\text{cl}}{4!} \sum_{\sum_{i=1}^4 n_i=0} q_{n_1}q_{n_2}q_{n_3}q_{n_4}-\frac{c_\text{cl}}{4!} q^4,\\\label{eq_intaction}
\end{eqnarray}
where we used the convention $q_0\equiv q$ and the definition $c=\beta c_\text{cl}$: taking the large temperature limit the classical interaction $c_\text{cl}$ is kept fixed.
Physical intuition tells us that in the semiclassical limit the contribution of the quantum fluctuations $q_{n\ne 0}$ must become negligible. Indeed, for small values of $\beta$ the coefficients in $\mathcal{S}_\text{free}$ diverge, as a consequence the modes $q_{n\ne 0}$ are pinned to zero. At large but finite temperatures the contributions of $q_{n\ne 0}$ are small and a perturbative expansion of the interacting action $\mathcal{S}_{\text{int}}$ around the free part $\mathcal{S}_\text{free}$ can be attempted.
From this perspective, let us define the free propagator of the nonzero Matsubara modes
\be\label{eq_prop_free}
\langle q_n^\dagger q_{n'}\rangle_\text{free}=\delta_{n,n'}\frac{\beta^2}{(2\pi n)^2+\beta^2}\, .
\ee
Hereafter, $\langle...\rangle_{\text{free}}$ means that the expectation values of the $q_{n\ne 0}$ Matsubara components with respect to the action $\mathcal{S}_\text{free}$ are taken: the modes $q_{n\ne 0}$ are treated as Gaussian variables with zero mean and variance Eq. \eqref{eq_prop_free}. Within this averaging procedure $q\equiv q_0$ is treated as a fixed parameter.
Hence, one can write
\be\label{eq_free_exp}
e^{-\mathcal{S}_\text{cl}}\int \dd q_{n\ne 0}\, \mathcal{O}e^{-\mathcal{S}_\text{free}-\mathcal{S}_\text{int}}\propto \langle \mathcal{O}e^{-\mathcal{S}_\text{int}} \rangle_{\text{free}}e^{-\mathcal{S}_\text{cl}}\, .
\ee
In the above an overall $q-$independent proportionality constant has been neglected, which can be fixed later on by imposing the correct normalization of Eq. \eqref{eq_rencl}, and the variables $q,q_n$ have been left out for the sake of lighter notation.
Following the standard perturbation theory procedure the right-hand side (r.h.s.) of Eq. \eqref{eq_free_exp} can be conveniently expressed in terms of connected expectation values. For example, given two observables $A$ and $B$, their connected expectation value is $\langle A B\rangle^\text{c}=\langle A B\rangle-\langle A\rangle \langle B\rangle$.
The r.h.s. of Eq. \eqref{eq_free_exp} can be rewritten in terms of the connected expectation values as
\be
\langle \mathcal{O}e^{-\mathcal{S}_\text{int}} \rangle_{\text{free}}e^{-\mathcal{S}_\text{cl}}=\mathcal{O}_\text{eff} e^{-\mathcal{S}_\text{eff}}
\ee
with
\begin{eqnarray}\label{eq_anh_eff1}
&&\mathcal{S}_\text{eff}=\mathcal{S}_\text{cl}-\sum_{j=1}^\infty \frac{(-1)^j}{j!}\langle (\mathcal{S}_\text{int})^j\rangle_{\text{free}}^\text{c}\, ,\\
&&\mathcal{O}_\text{eff}=\sum_{j=0}^\infty \frac{(-1)^j}{j!}\langle \mathcal{O} (\mathcal{S}_\text{int})^j\rangle_{\text{free}}^\text{c}\label{eq_anh_eff2}\, .
\end{eqnarray}

So far no approximations have been made and the true quantum expectation value can be recovered from Eq. \eqref{eq_rencl}, provided the exact expressions for $\mathcal{S}_\text{eff}$ and $\mathcal{O}_\text{eff}$ are available. Therefore, the exact computation of the effective action and observable is as hard as the original quantum problem and a proper truncation scheme is needed.
The small parameter of this perturbative expansion is not the classical interaction $c_\text{cl}$ itself, which can be arbitrary large, but rather the propagators of the $n\neq0$ Matsubara modes \eqref{eq_prop_free}: for $\beta\to 0$, the propagators are suppressed at least as $\beta^{2}$.
Therefore, the semiclassical expansion is organized in terms of how many propagators of the nonzero Matsubara modes  \eqref{eq_prop_free} are used: the $N^\text{th}$ order of the expansion is the sum of all the different terms with $N$ propagators. 
It can be easily understood that this truncation scheme requires the computation of only a finite number of terms: looking at $\mathcal{S}_\text{int}$ \eqref{eq_intaction}, in each term of the summation at least two nonzero Matsubara modes are present. Therefore, if we now consider the expansion of $\mathcal{S}_\text{eff}$ Eq. \eqref{eq_anh_eff1} and focus on the $j^\text{th}$ term, we can conclude this contributes with at least $j$ nonzero Matsubara propagators. Hence, selecting the $N^\text{th}$ term in the semiclassical expansion requires one to keep only terms up to  $j\le N$ in Eq. \eqref{eq_anh_eff1}. However, we stress once again that the semiclassical expansion is not equivalent to a simple truncation of Eq. \eqref{eq_anh_eff1}.
This approach is not a perturbative expansion in the interaction strength: albeit the $q_{n\ne 0}$ modes are only weakly coupled, the classical mode remains strongly interacting.
Following this general treatment, we frame the resulting expansion within a set of Feynman diagrams, compute explicitly the first corrections to natural observables, and compare the semiclassical expansion with the first-principles quantum results.
\subsection{Quantum corrections to the classical approximation}
\label{sec_anh_Fey}
\begin{figure}[t!]
\includegraphics[width=0.7\columnwidth]{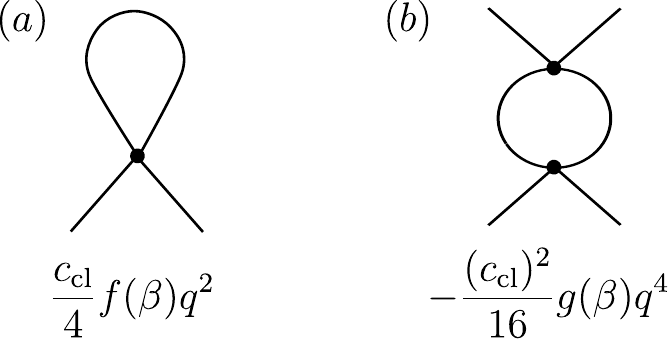}
\caption{\label{fig_feynman_1}
The diagrammatic representation of $\mathcal{S}_{\text{eff}}$ Eq. \eqref{eq_S_eff_2ord} up to the second order in the quantum corrections. Diagram $(a)$ takes into account first-order quantum corrections (notice the presence of a single propagator for the nonzero Matsubara modes), and diagram $(b)$ describes the second-order correction (two propagators are present). The functions $f(\beta)$ and $g(\beta)$ are defined in Eqs. \eqref{eq_f_def} and \eqref{eq_g_def}, respectively.
}
\end{figure}

The semiclassical expansion can be efficiently presented in terms of Feynman diagrams. We start by discussing the computation of the effective action $\mathcal{S}_\text{eff}$ \eqref{eq_anh_eff1}, then the generalization to $\mathcal{O}_\text{eff}$ will become clear.
First, we associate a vertex with four departing legs with the interaction $\mathcal{S}_{\text{int}}$, the legs represent the $q_n$ variables. Each vertex carries a factor $c_\text{cl}$ and a conservation law for the Matsubara frequencies, as is clear from Eq. \eqref{eq_intaction}. At each vertex there are at least two legs associated with nonzero Matsubara frequencies, and the computation of $\frac{1}{j!}\langle (\mathcal{S}_{\text{int}})^j\rangle_\text{free}^\text{c}$ goes as follows. Draw $j$ vertices. The propagators $\langle q^\dagger_n q_n\rangle_\text{free}$ are represented contracting the external legs of two interaction vertices. Since we are considering the connected parts of the correlators, the legs must be contracted in such a way that the final diagram does not have disconnected parts. Finally the sum over all the allowed Matsubara frequencies is performed. Non-contracted legs are generally present and associated with $q\equiv q_0$. 
There is a symmetry factor associated with this diagram which equals the number of permutations of legs and vertices that leave the diagram unchanged.
In Fig. \ref{fig_feynman_1}, we show the Feynman diagrams contributing to the effective action $\mathcal{S}_\text{eff}$ up to the second order in the semiclassical expansion, which result in the expression
\be\label{eq_S_eff_2ord}
\mathcal{S}_\text{eff}(q)=\Bigg[\frac{1}{2}q^2+\frac{c_\text{cl}}{4!} q^4\Bigg]+\Bigg[\frac{c_\text{cl}}{4}q^2 f(\beta)\Bigg]-\Bigg[\frac{c_\text{cl}^2}{16}g(\beta) q^4\Bigg]+...
\ee
where each term in brackets is a further order in the semiclassical expansion (from left to right: the classical approximation, the first-order, and the second-order quantum corrections).
The auxiliary functions $f(\beta)$ and $g(\beta)$ are defined as
\begin{eqnarray}\label{eq_f_def}
&&f(\beta)=\sum_{n\ne 0}\frac{\beta^2}{(2\pi n)^2+\beta^2}=\frac{1}{2}\left(\beta \coth(\beta/2)-2\right)\\
&&\nonumber
g(\beta)=\sum_{n\ne 0}\frac{\beta^4}{[(2\pi n)^2+\beta^2]^2}=\frac{4+\beta^2-4 \cosh\beta+\beta \sinh \beta}{8\sinh^2(\beta/2)}\\ &&\label{eq_g_def}
\end{eqnarray}
and represent the contributions of the loops in Fig. \ref{fig_feynman_1}.

For what concerns the observables, a natural choice is looking at the moments of the operator $\hat{q}$, hence we define
\be\label{eq_anh_Ol}
\mathcal{O}^{(\ell)}[\sqrt{\beta} \hat{q}]\equiv \frac{\beta^\ell}{(2\ell)!} \hat{q}^{2\ell}\, .
\ee

\begin{figure}[t!]
\includegraphics[width=1\columnwidth]{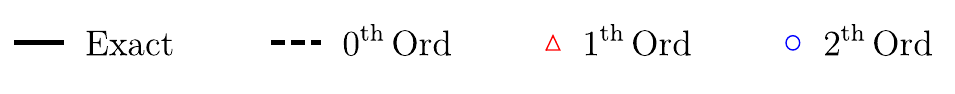}\\
\includegraphics[width=0.49\columnwidth]{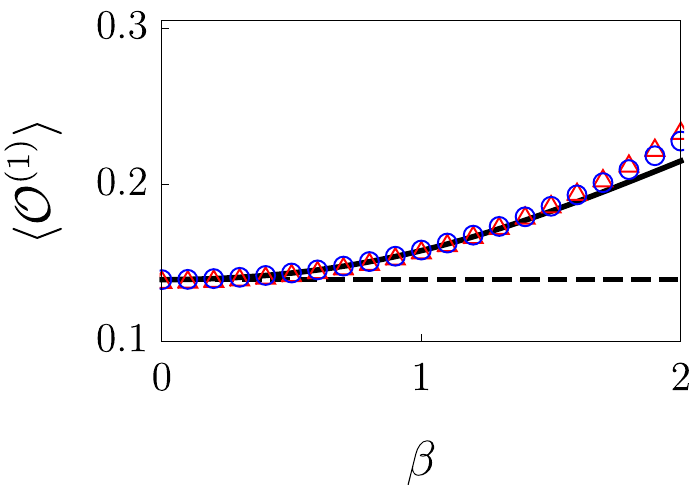}
\includegraphics[width=0.49\columnwidth]{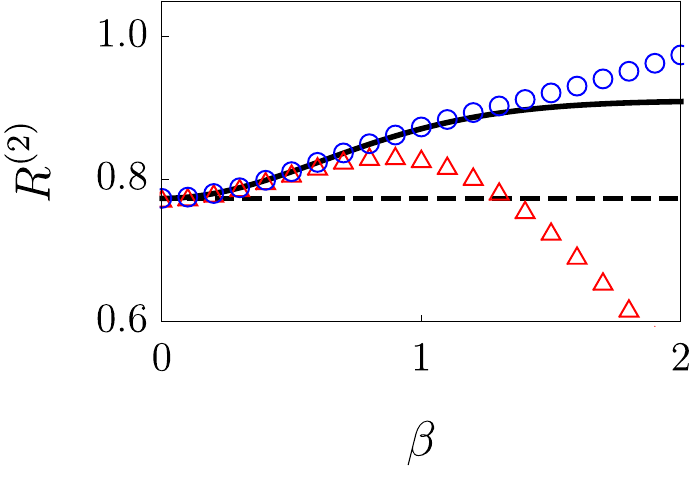}\\
\includegraphics[width=0.49\columnwidth]{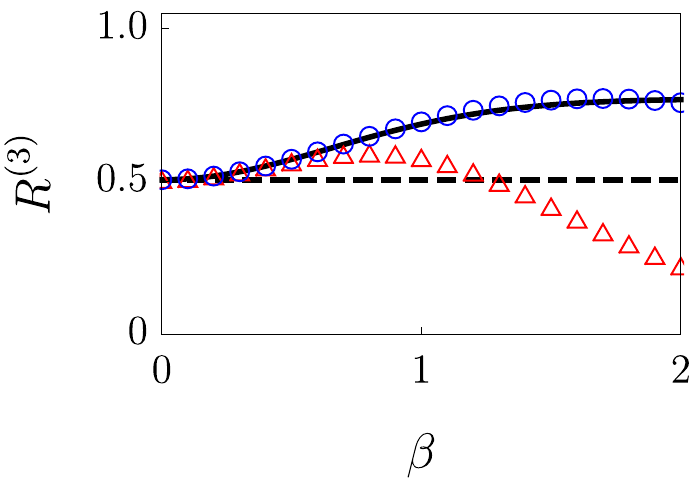}
\includegraphics[width=0.49\columnwidth]{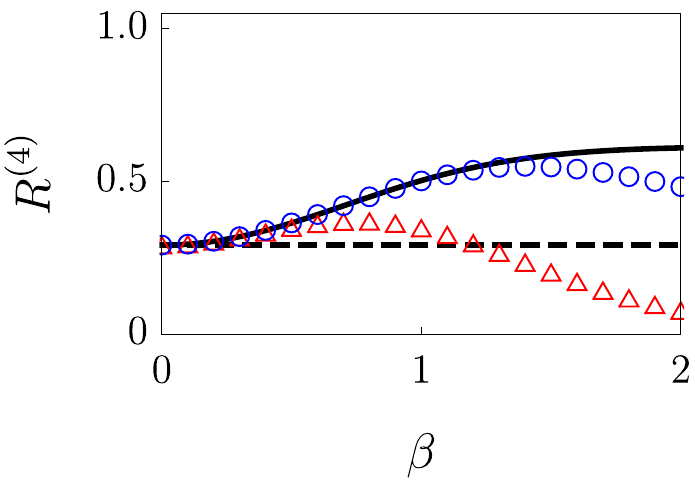}
\caption{\label{fig_Oncomparison_anh}
Comparisons of the semiclassical expansion results with the exact expectation values of the observables Eq. \eqref{eq_anh_Ol} for the anharmonic oscillator, numerically computed by discretizing the quantum Hamiltonian Eq. \eqref{eq_anhq}.
We chose the quantum interaction as $c=c_\text{cl}\beta$ with $c_\text{cl}=4!$, then the results are plotted as functions of the inverse temperature. We plot the ratios $R^{(\ell)}=\frac{1}{\ell!}\langle\mathcal{O}^{(\ell)\rangle}/(\langle\mathcal{O}\rangle)^\ell$ in order to stress the role of the interactions: for $c_\text{cl}=0$, the thermal ensemble is Gaussian, resulting in $R^{(\ell)}=1$. The fact that $R^{(\ell)}$ is far from unity confirms the strongly correlated nature of the system.
The $N^\text{th}$ order is obtained by computing $\langle \mathcal{O}^{(\ell)}\rangle$ to the required order in the quantum fluctuations, then the proper ratio of the observables is considered.
}
\end{figure}

The factorial is chosen for symmetry reasons: the Feynman rules for computing $\mathcal{O}^{(\ell)}_\text{eff}$ are simple generalizations of those for $\mathcal{S}_\text{eff}$ with the inclusion of a new interaction vertex associated with $\mathcal{O}^{(\ell)}$.
Up to the second order in the quantum corrections one finds
\begin{multline}
\mathcal{O}^{(\ell)}_\text{eff}[q]=\Bigg[\frac{q^{2\ell}}{(2\ell)!}\Bigg]+
\Bigg[\frac{f(\beta) q^{2(\ell-1)}}{2[2(\ell-1)]!}\Bigg]\\
+\Bigg[ \frac{ f^2(\beta)q^{2(\ell-2)}}{8[2(\ell-2)]!}-\frac{1}{4}\frac{c_\text{cl}g(\beta)q^{2\ell}}{[2(\ell-1)]!}\Bigg]+...\,.
\end{multline}
As for Eq. \eqref{eq_S_eff_2ord}, each bracket represents a further order in the semiclassical expansion.
Above, we use the convention that $q-$powers with negative exponents are actually absent: hence, for $\ell=1$ the term $\propto q^{2(\ell-1)}$ must be discarded.
In Fig. \ref{fig_Oncomparison_anh}, the semiclassical expansions for the expectation values of $\mathcal{O}^{(\ell)}$ are benchmarked with the numerically exact diagonalization of the quantum problem Eq. \eqref{eq_anhq}, showing a good agreement.

\begin{figure*}[t!]
\begin{center}
\includegraphics[width=0.9\columnwidth]{anh_obs_legend.pdf}\\
\includegraphics[width=0.3\textwidth]{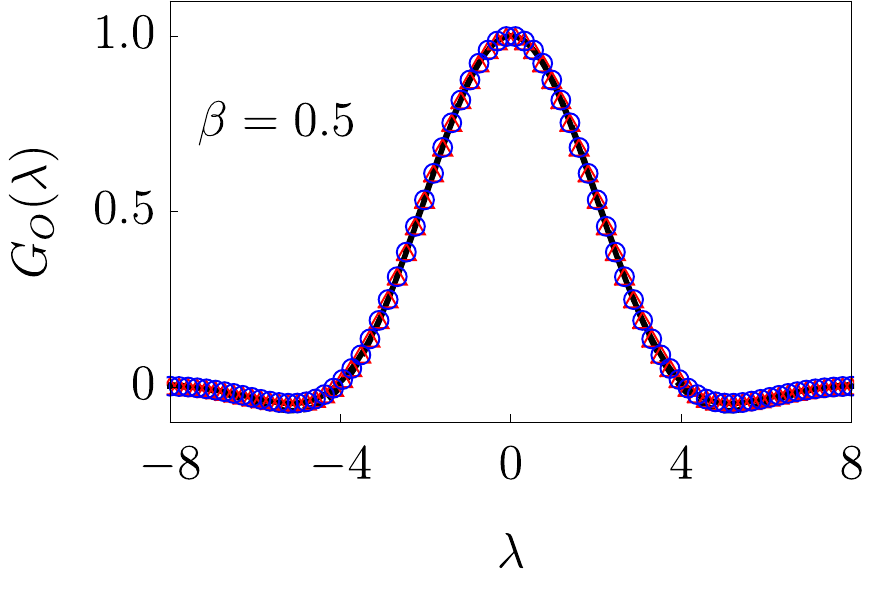}\hspace{1pc}
\includegraphics[width=0.3\textwidth]{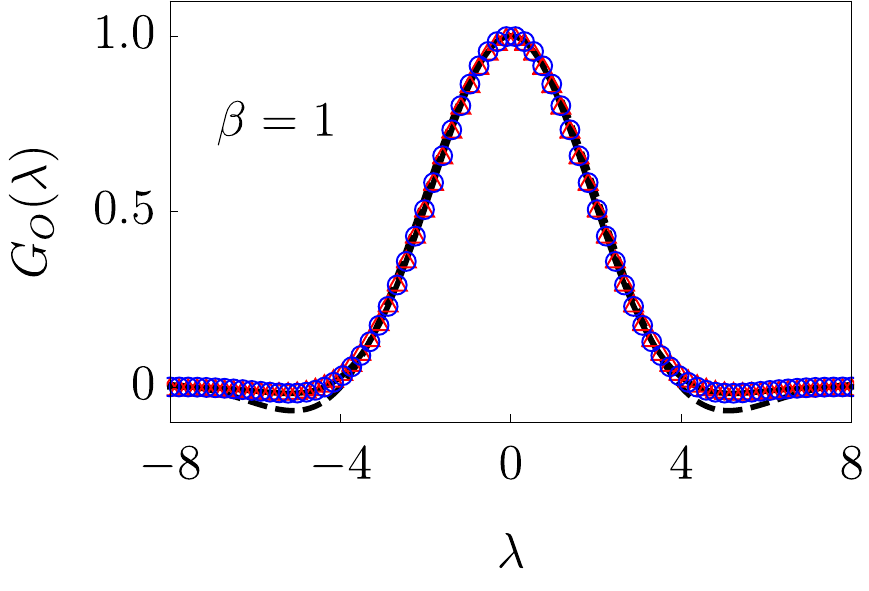}\hspace{1pc}
\includegraphics[width=0.3\textwidth]{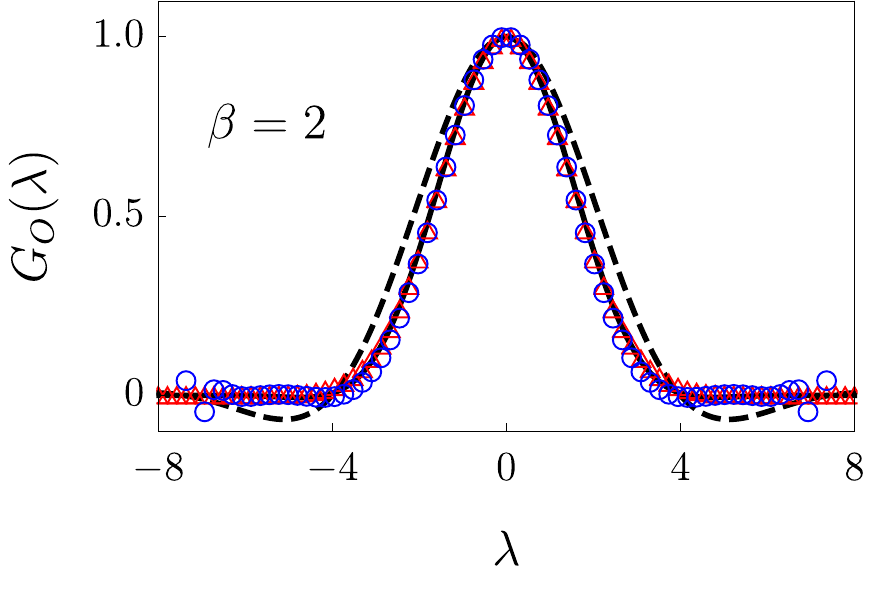}
\end{center}
\caption{\label{fig_fcscomparison_anh}
We consider the generating function $G_O(\lambda)$ of the FCS of the observable $O=\sqrt{\beta}\hat{q}$ for different temperatures, comparing the exact numerically computed values to results of the semiclassical approximation [Eqs. \eqref{eq_exp_gen} and \eqref{eq_gen_2ord}].
At high temperatures $\beta=0.5$, the agreement is excellent already with the classical prediction. The interaction $c=c_\text{cl}\beta$ is chosen as in Fig. \ref{fig_Oncomparison_anh}, i.e., $c_\text{cl}=4!$. Decreasing the temperature the classical approximation departs from the quantum value, which is still well described by the next orders of the expansion.
Notice that the discrepancy starts to appear from the tails of the distribution: indeed, as we discuss in the main text, the semiclassical approximation of the generating function has a natural cutoff $|\lambda|\lesssim \beta^{-1}$ beyond which it cannot be applied any longer. Increasing $\beta$ the cutoff is decreased and corrections become more pronounced.
}
\end{figure*}

In addition to the moments of $\hat{q}$, the semiclassical approximation can be used to study its FCS. 
The FCS is the full probability distribution of measuring a certain value for an observable $O[\sqrt{\beta} \hat{q}]$. For example, in the following, we focus on the observable $O=\sqrt{\beta} \hat{q}$ (the rescaling by $\sqrt{\beta}$ is introduced for later convenience), its FCS being the probability of finding the particle at position $q$. 
More formally, we are interested in the probability $P_O(w)$ defined as

\be
P_O(w)\equiv\frac{1}{\mathcal{Z}_q}\text{Tr}[ \delta(w-O)e^{-\beta \hat{H}}]\equiv\langle \delta(w-O)\rangle\, .
\ee
The distribution above is not directly amenable to the semiclassical analysis, hence we rather define its generating function $G_O(\lambda)$
\be\label{eq_gen_fcs}
G_O(\lambda)=\langle e^{i\lambda O}\rangle\, ,\hspace{1pc}P_O(w)=\int \frac{\dd \lambda}{2\pi}e^{-iw \lambda}G_O(\lambda)\, .
\ee
The generating function in the form Eq. \eqref{eq_gen_fcs} can be treated within our semiclassical approach since we can set $\mathcal{O}=e^{i\lambda O}$ and evaluate this effective observable by means of Eq. \eqref{eq_anh_eff2}. Moreover, due to the exponential form of the observable $\mathcal{O}$ the series defining $\mathcal{O}_\text{eff}$ Eq. \eqref{eq_anh_eff2} can be further resummed into a more convenient expression. Indeed, one can establish the following identity:
\begin{multline}
\mathcal{O}_\text{eff}e^{-\mathcal{S}_\text{eff}}=e^{-\mathcal{S}_\text{eff}}\sum_{j=0}^\infty \frac{(-1)^j}{j!}\langle e^{i\lambda O} (\mathcal{S}_\text{int})^j\rangle_{\text{free}}^\text{c}\\
=\exp\Big[\sum_{j=1}^\infty\frac{1}{j!}\langle(i\lambda O-\mathcal{S}_\text{int})^j \rangle_\text{free}^\text{c}\Big]\, .
\end{multline}

The exponentiation of the above series leads to the natural definition of the effective action  
\be\label{eq_eff_Olambda}
\mathcal{S}_{\text{eff}}^{O,\lambda}[q]=-\sum_{j=1}^\infty\frac{1}{j!}\langle(i\lambda O-\mathcal{S}_\text{int})^j \rangle_\text{free}^\text{c}\, ,
\ee
which depends on the observable $O$ and the parameter $\lambda$.
Above, we stress that $\mathcal{S}_{\text{eff}}^{O,\lambda}[q]$ is a function of $q$, the classical ($n=0$) field.
With this definition the generating function is expressed as
\be\label{eq_exp_gen}
G_O(\lambda)=\frac{\int \dd q\, e^{-\mathcal{S}^{O,\lambda}_\text{eff}[q]}}{\int \dd q\, e^{-\mathcal{S}_\text{eff}[q]}}\, .
\ee

Equation \eqref{eq_exp_gen} is in principle exact: we performed the semiclassical expansion approximating Eq. \eqref{eq_eff_Olambda} with the insertion of a finite number of nonzero Matsubara propagators \eqref{eq_prop_free}. More precisely, the $N^\text{th}$ order of the expansion contains all those terms with $N$ propagators of the $n\neq0$ Matsubara modes. At any order of the expansion the sum in Eq. \eqref{eq_eff_Olambda} is finite with only a finite number of terms to be computed.
A word of caution is in order regarding the behavior of the expansion for arbitrary values of $\lambda$: even though the expansion induced by truncating Eq. \eqref{eq_eff_Olambda} is not perturbative in $\lambda$ (the interacting action of the classical mode is taken into account exactly), any approximated truncation of Eq.  \eqref{eq_eff_Olambda} will eventually fail if $\lambda$ is too large.
As an example, let us consider $O[\sqrt{\beta} \hat{q}]=\sqrt{\beta} \hat{q}$ which, in the path integral language and after the replacement Eq. \eqref{eq_E_mat}, states $i\lambda O=i\lambda \sum_n q_n$. Since all the nonzero Matsubara frequencies are suppressed as $q_{n\ne 0}\sim \beta$, as it is clear from the propagator \eqref{eq_prop_free}, this sets a natural scale $|\lambda|\le\Delta^{-1}$ with $\Delta\ll \beta$ in which we can reasonably trust the semiclassical expansion of the effective action.
With the ultimate goal of accessing the FCS of the observable we define a $\Delta-$regularized FCS $P_O^\Delta$ by imposing a hard cutoff in the integral Eq. \eqref{eq_gen_fcs}. This is equivalent to the calculation of the coarse grained FCS defined as 
\begin{multline}\label{eq_cg_fcs}
P_O^\Delta(w)=\int_{-\Delta^{-1}}^{\Delta^{-1}} \frac{\dd \lambda}{2\pi}e^{-iw \lambda}G_O(\lambda)\\
=\int \frac{\dd w'}{\pi} \frac{\sin(\Delta^{-1}(w-w'))}{w-w'}P_O(w')\, .
\end{multline}
Hence, rather than directly accessing the FCS, the semiclassical approximation yields a coarse-grained version of it.

The quality of the semiclassical expansion is estimated by looking at the typical lengthscale on which the generating function $G_O(\lambda)$ decays: if for $|\lambda|\gtrsim \Delta^{-1}$ $G_O(\lambda)$ is negligibly small, the semiclassical approximation is expected to be good.
As a benchmark we compute $\mathcal{S}_{\text{eff}}^{O,\lambda}$ for the operator $O[\sqrt{\beta} \hat{q}]=\sqrt{\beta} \hat{q}$ up to the second order in the quantum corrections
\begin{multline}\label{eq_gen_2ord}
\mathcal{S}_{\text{eff}}^{O,\lambda}[q]-\mathcal{S}_{\text{eff}}[q]\equiv\\\Bigg[-i\lambda q\Bigg]+\Bigg[-\frac{\lambda^2f(\beta)}{2}\Bigg]+
\Bigg[-c_\text{cl} \frac{\lambda^2 g(\beta)}{4}q^2\Bigg]...\, .
\end{multline}
As before, each pair of square brackets corresponds to the next order of the expansion. For consistency, $\mathcal{S}_{\text{eff}}[q]$ must be expanded to the same order, see Eq.  \eqref{eq_S_eff_2ord}.
In Fig. \ref{fig_fcscomparison_anh}, we compare the semiclassical expansion for the generating function and the resulting FCS of the position of the particle in the anharmonic potential with the exact first-principles results.

\section{The 1d interacting Bose gas}
\label{sec_Bosegas}

Having presented the method in the simple case of a single degree of freedom, we now turn to the main purpose of our investigation, namely describing many-body quantum systems.
For the sake of concreteness and, later, benchmarking, we focus on the Lieb-Liniger (LL) model describing 1D bosons with contact pairwise interactions. In the language of the second quantization, the Hamiltonian is
\be\label{eq_LL_H}
\hat{H}=\int \dd x \left\{\frac{1}{2m} \partial_x\hat{\psi}^\dagger\partial_x\hat{\psi} +c\hat{\psi}^\dagger\hat{\psi}^\dagger \hat{\psi}\hat{\psi} -\mu \hat{\psi}^\dagger\hat{\psi}\right\}\, ,
\ee
where $[\hat{\psi}(x),\hat{\psi}^\dagger(y)]=\delta(x-y)$.
Hereafter we consider only the repulsive case, $c>0$, since the attractive phase is unstable in thermal equilibrium \ocite{doi:10.1063/1.1704156,PhysRevA.11.265}.

As it was mentioned before, this model is integrable \ocite{PhysRev.130.1605,PhysRev.130.1616}.
Because of integrability its thermodynamics can be solved exactly by means of the thermodynamic Bethe ansatz (TBA) \ocite{takahashi2005thermodynamics}, and the expectation values of certain observables can be computed exactly. Focusing on the Lieb-Liniger model provides an important benchmark for our approach, however we emphasize that it is fully general and does not rely on integrability at all.
For example, the moments of the density operator $\langle [\hat{\psi}^\dagger(x)]^n [\hat{\psi}(x)]^n\rangle$ in thermal states have been evaluated exactly in Refs. \ocite{PhysRevLett.120.190601,Bastianello_2018} for arbitrary integer $n$ (see also Refs. \ocite{PhysRevLett.107.230405,Pozsgay_2011} for previous results for $n\le 4$).
The semiclassical expansion, in addition to being able to correctly reproduce known results of the LL model, also provides new insights into quantities that are not accessible with the state-of-the-art techniques in integrability, the prime example being the two-point correlator $\langle \hat{\psi}^\dagger(x)\hat{\psi}(y)\rangle$, also known as the single-particle density matrix, which is routinely measured in the laboratory through the time-of-flight capture of the momentum-density distribution \ocite{RevModPhys.80.885}.
Another quantity of interest we consider is the FCS of the number of particles on an interval, which has recently been studied in the classical limit by two of us \ocite{PhysRevLett.122.120401}. In the following, we will first present a short summary of the integrability of the LL model and quote exact results that will be used for benchmarking before proceeding to the discussion of the semiclassical expansion.

\subsection{The integrability of the Lieb-Liniger model}
\label{sec_integrability}
The integrability of the LL model was established in the original works of Lieb and Liniger \ocite{PhysRev.130.1605,PhysRev.130.1616} which found exact eigenstates of the model. The eigenstates can be labeled by collections of quantum numbers $|\{k\}_{i=1}^N\rangle$, called rapidities or quasimomenta, which generalize the momentum modes of free bosons/fermions to the interacting case. The set of quasimomenta can be interpreted as a collection of excitations which undergo pairwise scatterings, hence the set $\{k\}_{i=1}^N$ is left unscathed by the time evolution. Despite being elastic, the scattering processes are nontrivial and the interactions are encoded in the two-body scattering matrix, which in this case is just a complex number $S(k)=(k-2i m c)/(k+2imc)$.
Within the thermodynamic limit, one adopts a coarse-grained description in terms of a
filling function $\vartheta(k)$, which generalizes the mode-density occupation number
of free systems: this is the founding idea of the thermodynamic Bethe ansatz \ocite{takahashi2005thermodynamics}. A proper introduction to the TBA and,
more generally, integrability goes beyond the scope of the current work, therefore we simply quote the results of interest for us. The interested reader can refer to Refs. \ocite{korepin1997quantum,takahashi2005thermodynamics}  for further details.

The filling function $\vartheta(k)$ associated with a given thermal state is explicitly determined through the following integral equation:
\be\label{eq_TBA}
\log\frac{1-\vartheta(k)}{\vartheta(k)}=\beta \left[\frac{k^2}{2m} -\mu\right]+\int \frac{\dd q}{2\pi} \varphi(k-q)\log(1-\vartheta(k))\, ,
\ee
where  $\varphi(k)=-i\partial_k S(k)$. In general, no analytic solution to
this equation is known, but it can be easily solved numerically.
Once the filling function is determined, this in principle fixes any local property of the system, as well as the correlation functions, however actually evaluating them is a hard task on its own.

Very recently the expectation values of all the density moments $\langle(\hat{\psi}^\dagger)^n(\hat{\psi})^n\rangle$ for arbitrary $n$ have been computed \ocite{PhysRevLett.120.190601,Bastianello_2018}, the results being expressed in terms of a set of coupled integral equations as reported below.
The density moments are accessed by expanding the following generating function in the dummy variable $Y$ around zero,
\be\label{eq_quantum_GF}
1+\sum_{n=1}^\infty Y^n \frac{2^n (2mc)^n}{(n!)^2}\langle (\hat{\psi}^\dagger)^n(\hat{\psi})^n\rangle\,=\,\exp\left(\frac{1}{\pi }\sum_{n=1}^\infty Y^n \mathcal{G}_n\right)\, ,
\ee
where
\be
\mathcal{G}_n=\frac{(2m c)^{2n-1} }{n}\int\dd k\, \vartheta(k) \xi^\text{dr}_{2n-1}(k)\, .
\ee
The dressing operation $\xi\to \xi^{\text{dr}}$ is defined as the solution of the following linear integral equation
\be
\xi^\text{dr}_{n}(k)=\xi_{n}(k)+\int \frac{\dd q}{2\pi} \varphi(k-q)\vartheta(q)\xi^\text{dr}_{n}(q)\,.
\ee
The auxiliary functions $\xi_n(k)$ are obtained recursively by solving the following set of integral equations [below we define $\Gamma(k)=k (2m c)^{-1}\varphi(k)$]:
\begin{multline}\label{int_q1}
\xi_{2n}(k)=\int \frac{\dd p}{2\pi} \vartheta(p)\Big\{\Gamma(k-p)[2\xi^\text{dr}_{2n-1}(p)\\
-\xi^{\text{dr}}_{2n-3}(p)]-\varphi(k-p)\xi^\text{dr}_{2n-2}(p)\Big\},
\end{multline}
\begin{multline}\label{int_q2}
\xi_{2n+1}(k)=\delta_{n,0}+\int \frac{\dd p}{2\pi} \vartheta(p)\Big\{\Gamma(k-p)\xi^\text{dr}_{2n}(p)\\
-\varphi(k-p)\xi^\text{dr}_{2n-1}(p)\Big\}\, ,
\end{multline}
where $\xi_{n<0}=0$.
The integral equations have a recursive structure, since the equation for $\xi_n$ requires the knowledge of $\xi_{n'\le n}$: these can be solved finding the function $\xi_0(\lambda)$, then the integer $n$ is progressively increased and the functions $\xi_n(\lambda)$ are recursively determined. Note that extracting the $n^\text{th}$ density moment requires finding the first $2n-1$ auxiliary functions $\xi_n$.

\subsection{The semiclassical expansion}
\label{sec_LLsemi}

The semiclassical expansion of the 1D interacting Bose gas closely follows the study of the anharmonic oscillator presented in Sec. \ref{sec_singleboson}, therefore we limit the discussion to the key points of the analysis.
We start with a path integral representation of the thermal state and introduce a complex-valued field $\psi_E(\tau,x)$ governed by the action
\begin{multline}\label{Euclidean_action}
\mathcal{S}_E=\int_0^\beta \dd \tau \int \dd x\,\Big\{\frac{1}{2}(\psi_E^*\partial_\tau\psi_E-\psi_E\partial_\tau\psi_E^*)\\
+\frac{1}{2m}|\partial_x\psi_E|^2
-\mu|\psi_E|^2+c|\psi_E|^4\Big\}\, ,
\end{multline}
with periodic boundary conditions in the Euclidean time $\tau\in [0,\beta]$. Similarly to the anharmonic oscillator, the field is split into the Matsubara modes,
\be
\psi_E(\tau,x)=\frac{\psi(x)}{\sqrt{\beta}}+\sum_{n\ne 0} e^{i 2\pi n \tau/\beta}\frac{\psi_n(x)}{\sqrt{\beta}}\, .
\ee
Since $\psi_E$ is complex, the modes $\psi_n$ are independent complex fields: this is in contrast to the anharmonic oscillator case.
The field $\psi(x)$ can be identified as the classical variable, the action of which is renormalized by the integration over the nonzero frequency Matsubara modes. To this end, we split the action into three parts $\mathcal{S}_E=\mathcal{S}_\text{cl}+\mathcal{S}_\text{free}+\mathcal{S}_\text{int}$
\begin{eqnarray}
&&\mathcal{S}_\text{cl}[\psi]=\int \dd x \left\{\frac{|\partial_x\psi|^2}{2m}+c_\text{cl}|\psi|^4-\mu |\psi|^2\right\}\, ,\\
&&\nonumber\mathcal{S}_\text{free}[\psi_{n\ne 0}]=\int \dd x\, \Bigg\{\sum_n  [i2\pi n\beta^{-1}\\
&&+(4c_\text{cl} d-\mu)]|\psi_n|^2+\frac{| \partial_x\psi_n|^2}{2m}\Bigg\}\label{eq_LL_Sfree}\,,\\
&&\nonumber\,\\
&&\nonumber\mathcal{S}_\text{int}[\psi,\psi_{n\ne 0}]=\int \dd x\,\Bigg\{\sum_{{\scriptsize\begin{matrix}n_1+n_2=\\n_3+n_4\end{matrix}}} c_\text{cl}\psi^*_{n_1}\psi^*_{n_2}\psi_{n_3}\psi_{n_4}\\
&&
-c_\text{cl}|\psi|^4+4 c_\text{cl}d\sum_{n\ne 0}|\psi_n|^2\Bigg\}
\label{eq_LL_S_int}
\end{eqnarray}
\begin{figure*}[t!]
\begin{center}
\includegraphics[width=0.9\columnwidth]{anh_obs_legend.pdf}\\
\includegraphics[width=0.35\textwidth]{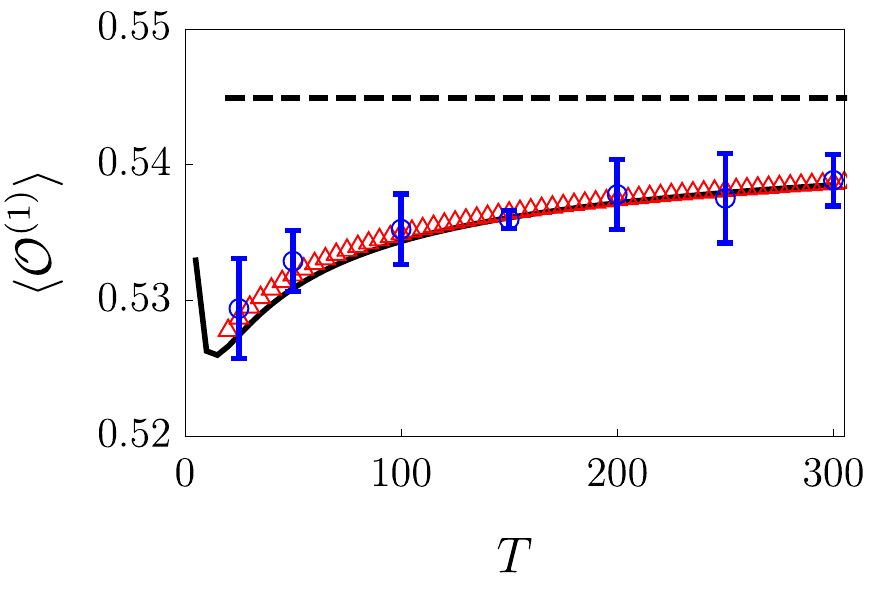}
\includegraphics[width=0.35\textwidth]{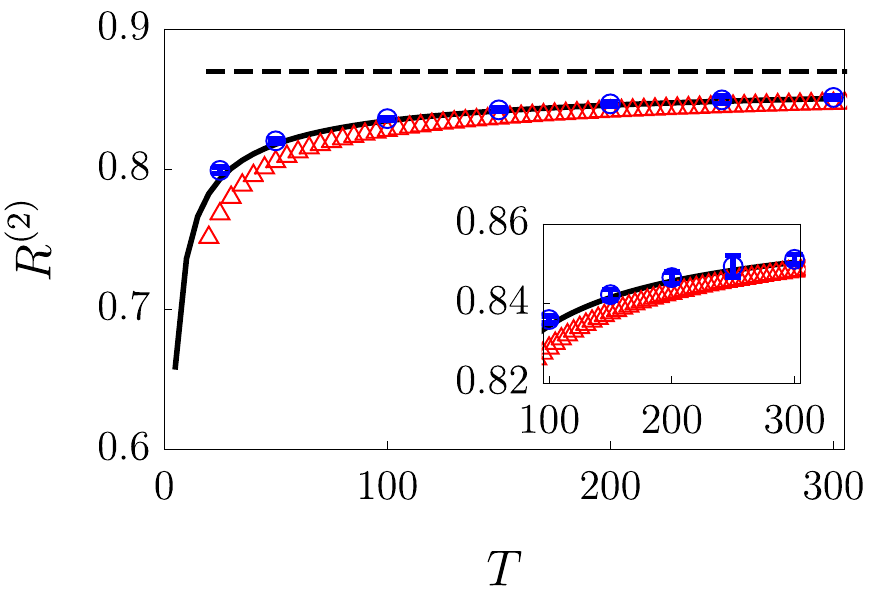}\\\includegraphics[width=0.35\textwidth]{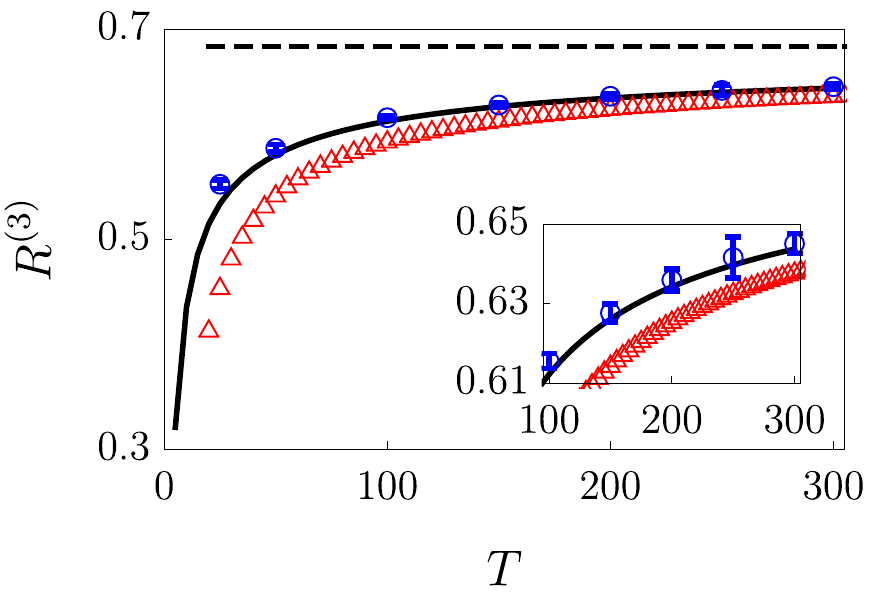}
\includegraphics[width=0.35\textwidth]{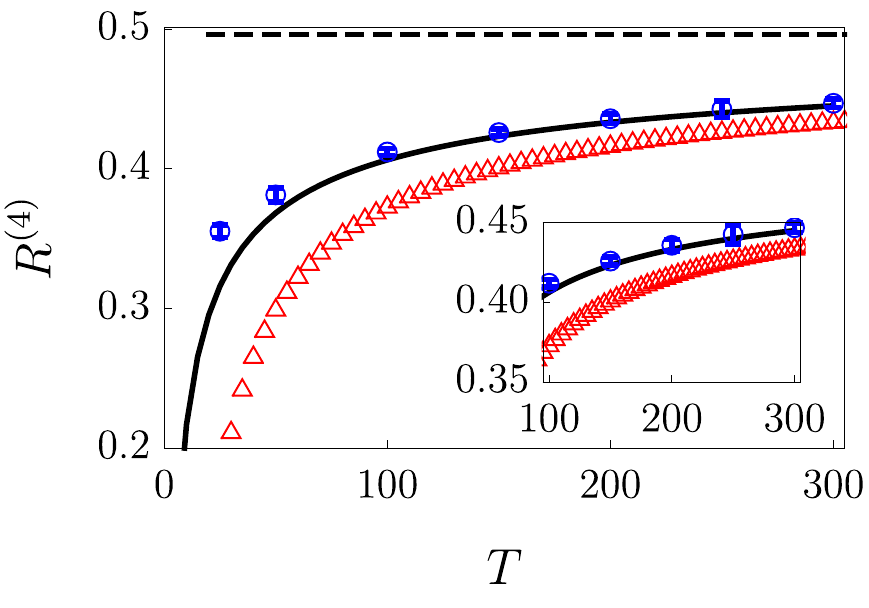}
\end{center}
\caption{\label{fig_LL_onept}Above we compare the semiclassical expansions of the expectation values of the moments of the local density with the exact results presented in Sec. \ref{sec_integrability}.
More specifically, we consider $\mathcal{O}^{(\ell)}[\sqrt{\beta} \hat{\psi}]=\beta^\ell [\hat{\psi}(x)]^\ell [\hat{\psi}(x)]^\ell$ and $R^{(\ell)}=\langle \mathcal{O}^{(\ell)}\rangle/[\ell! (\langle \mathcal{O}^{(1)}\rangle)^\ell]$. The normalization of $R^{(\ell)}$ is chosen in such a way that, in the absence of interactions, we would have had $R^{(\ell)}=1$: thus, any deviation from unity signals the strongly correlated nature of the model.
In the LL Hamiltonian \eqref{eq_LL_H}, we choose $m=1/2$, $c=c_\text{cl} \beta$ with $c_\text{cl}=1$ and $\mu=1$ and then vary $T=\beta^{-1}$. 
We see that the exact prediction (solid line) in the high-temperature limit clearly approaches the classical result (dashed line), but corrections are important and they are nicely captured by the first-order (red triangles) and second-order (blue circles) quantum corrections.
The classical results and the first-order corrections are computed with the transfer matrix method (see Appendix \ref{sec_TM}), while the second-order corrections are obtained with the Metropolis-Hastings algorithm (Appendix \ref{sec_metr}).
}
\end{figure*}
The classical interaction strength is defined as $c_\text{cl}=c\beta^{-1}$, same as before. The classical limit is attained by sending $\beta\to 0$ while keeping $c_\text{cl}$ and $\mu$ constant.
Indeed, each term in the interacting part of the action $\mathcal{S}_\text{int}$ contains at least two $n\neq0$ Matsubara modes and the prefactor in $\mathcal{S}_\text{free}$ diverges as $\beta\to 0$. Hence, in the high-temperature limit the modes $\psi_{n\ne 0}$ are pinned to zero and only the field $\psi(x)\equiv \psi_0(x)$ is free to fluctuate.
In the splitting of the action above we included a parameter $d$ which does not appear in the original action $\mathcal{S}_E$. Indeed, summing back the three terms of the action the $d-$dependece is canceled out.
Naively one could pose $d=0$ in such a way to simplify the interacting part of the action, but this choice is problematic when $\mu>0$: this would lead to an action $\mathcal{S}_\text{free}$ which is unbounded from below and cannot serve as a starting point of the perturbation theory any longer.
Indeed, whereas in the LL model both positive and negative values of $\mu$ are allowed, the noninteracting bosons are well-defined only for $\mu<0$ and the interactions play a crucial role in making the case of positive chemical potential a well-defined ensemble.
We choose the free parameter $d$ to be the density evaluated in the classical limit, $d=\langle |\psi|^2\rangle_{\text{cl}}$, thus neglecting the $n\neq0$ Matsubara modes. This choice is inspired by the Gaussian approximation of the interaction: keeping only up to two nonzero frequency Matsubara modes in the interacting term (which will dominate the high-temperature limit) one has 
\be
c_\text{cl}\sum_{n_1+n_2=n_3+n_4}\psi^*_{n_1}\psi^*_{n_2}\psi_{n_3}\psi_{n_4}\simeq c_\text{cl}|\psi|^4 +c_\text{cl}\sum_{n\ne 0} |\psi_n|^2 |\psi|^2.
\ee
Treating the above in perturbation theory under the assumption that $|\psi_n|^2$ is small, but first integrating out the classical mode, one replaces $|\psi|^2\to \langle|\psi|^2\rangle$, which in first approximation can be computed with the classical action. It is numerically verified that $4 c_\text{cl}d-\mu>0$, making $\mathcal{S}_\text{free}$ well defined, and we stress that the Gaussian approximation is only used to justify a convenient value for the parameter $d$, but then we apply our method to systematically include quantum corrections on top of the classical approximation.

The expectation values of quantum obervables $\mathcal{O}$ can now be computed in an effective one-dimensional classical field theory, similarly to how it was done for the quantum anharmonic oscillator,

\be\label{eq_eff_1d}
\frac{1}{\mathcal{Z}_{\text{q}}}\text{Tr}[\mathcal{O}[\sqrt{\beta}\hat{\psi}] e^{-\beta \hat{H}}]=\frac{1}{\mathcal{Z}}\int \mathcal{D} \psi\, \mathcal{O}_\text{eff}[\psi] e^{-\mathcal{S}_\text{eff}[\psi]}\, .
\ee
The effective action and observables are defined as per Eqs. \eqref{eq_anh_eff1} and \eqref{eq_anh_eff2} where $\mathcal{S}_\text{int}$ is now Eq. \eqref{eq_LL_S_int}, and the free expectation value are computed with respect to Eq. \eqref{eq_LL_Sfree}.
Therefore, the basic ingredient we need is the free propagator of the $n\neq0$ Matsubara modes,
\begin{multline}\label{eq_psi_free_c}
\langle \psi^*_n(x) \psi_{n'}(y)\rangle_\text{free}=
\delta_{n,n'}\int \frac{\dd k}{2\pi} \frac{e^{ik (x-y)}}{i2\pi n\beta^{-1} +k^2+\Omega}\\
=\delta_{n,n'}\frac{\exp\left[-|x-y|\sqrt{\Omega+i2\pi n \beta^{-1}}\right]}{2\sqrt{\Omega+i2\pi n \beta^{-1}}}\, ,
\end{multline}
where we defined $\Omega=4 c_\text{cl} d-\mu$.

When performing the quantum expansion we often need $\sum_{n\ne 0} \langle \psi_n^*(x)\psi_n(x)\rangle_\text{free}$. This quantity is ill-defined unless the point-splitting regularization is imposed: the limit of equal positions is taken only after the summation has been performed. For later convenience we define $C(x-y)=\sum_{n\ne 0}\langle \psi^*_n(x)\psi_n(y)\rangle_\text{free}$ which can be written as
\be\label{eq_C_f}
C(x-y)=
\int \frac{\dd k}{2\pi}\, e^{ik(x-y)}\left[\frac{\beta}{e^{\beta(k^2+\Omega)}-1}-\frac{1}{k^2+\Omega}\right]\, .
\ee
Notice that the limit $\lim_{x\to 0}C(x)$ is well defined.

We can now expand $\mathcal{S}_\text{eff}$ by including the quantum corrections. Analogously to what was done in Sec. \ref{sec_singleboson}, we organize the expansion in terms of the number of the propagators of the nonzero frequency Matsubara modes. The expansion can be efficiently described in terms of Feynman diagrams with minor modifications compared to those presented for the anharmonic oscillator.
The effective action up to the second order in the quantum fluctuations is readily computed,

\begin{multline}\label{eq_eff_S_LL}
\mathcal{S}_\text{eff}[\psi]=\Bigg[S_\text{cl}[\psi]\Bigg]+
\Bigg[4c_\text{cl}C(0)\int \dd x \,|\psi(x)|^2\Bigg]\\
+\Bigg[16d c_\text{cl}^2\int \dd y\, F(y)\int \dd x \, |\psi(x)|^2
\\
-8 c_\text{cl}^2 \int \dd x \dd y\,|\psi(x)|^2 |\psi(y)|^2F(x-y)\\
-2c_\text{cl}^2\Re\left(\int \dd x\dd y\,\psi^*(x)\psi^*(x)\psi(y)\psi(y)R(x-y)\right)\Bigg]+...
\end{multline}
Above we used the square brackets to point out the different orders in the expansion, as we already did in Sec. \ref{sec_singleboson}. The auxiliary functions $F(x)$ and $R(x)$ are defined as

\begin{multline}\label{eq_F_aux}
F(x)=\sum_{n\ne 0}\langle\psi^*_n(x)\psi_n(0)\rangle_\text{free}\langle\psi^*_n(0)\psi_n(x)\rangle_\text{free}\\
=\frac{1}{4}\sum_{n\ne  0}\frac{\exp\left[-2|x|\sqrt{\Omega+i2\pi n \beta^{-1}}\right]}{\Omega+i2\pi n \beta^{-1}}\, ,
\end{multline}

\begin{multline}\label{eq_G_aux}
R(x)=\sum_{n\ne 0}\langle \psi_n(x)\psi^*_n(0)\rangle_\text{free}\langle \psi_{-n}(x)\psi^*_{-n}(0)\rangle_\text{free}\\
=\frac{1}{4}\sum_{n\ne 0}\frac{\exp\left[-|x|2\Re\sqrt{\Omega+i2\pi n \beta^{-1}}\right]}{\sqrt{\Omega^2+(2\pi n \beta^{-1})^2}}\, .
\end{multline}
The effective action resulting from the semiclassical expansion is strictly local up to the first-order correction, which simply consists of a renormalization of the chemical potential. Classical expectation values in thermal ensembles with local actions can be efficiently computed with a transfer matrix approach (see App. \ref{sec_TM}) which transforms the classical field problem into a quantum system with a finite number of degrees of freedom.

Beyond the first-order terms the effective action develops nonlocal terms and the transfer matrix approach is not applicable any longer, hence we revert to the Metropolis-Hastings algorithm \ocite{doi:10.1080/00031305.1995.10476177} (see App. \ref{sec_metr}).
Long-range interactions are challenging, but this is not the case at hand: as can be seen from the auxiliary functions $F(x)$ Eq. \eqref{eq_F_aux} and $R(x)$ Eq. \eqref{eq_G_aux}, the nonlocal terms appearing in the effective action are short-ranged and do not undermine the efficiency of the algorithm. In particular, the range of the interaction is reduced as the temperature is increased.

Having computed the effective action, we now revert to the observables. In view of the integrability-based exact results presented in Sec. \ref{sec_integrability}, the moments of the density operator $\mathcal{O}^{(\ell)}[\sqrt{\beta}\hat{\psi}]=\beta^\ell [\hat{\psi}^\dagger(x)]^\ell[\hat{\psi}(x)]^\ell$ are natural candidates. The expansion up to the second-order is
\begin{multline}
\mathcal{O}_\text{eff}^{(\ell)}[\psi]=\Bigg[|\psi|^{2\ell}\Bigg]
+\Bigg[\ell^2 C(0) |\psi|^{\ell-1}\Bigg]\\
+\Bigg[\frac{\ell^2(\ell-1)^2 C^2(0)}{2} |\psi|^{2(\ell-2)}\\
-c_\text{cl}\ell (\ell-1)2\Re\left(|\psi|^{2(\ell-2)}(\psi)^2 \int \dd x\, R(x)\psi^*(x)\psi^*(x)\right)
\\
-4 \ell^2c_\text{cl}|\psi|^{2(\ell-1)}\int \dd x F(x)(|\psi(x)|^2-d)\Bigg]+...
\end{multline}

For the sake of brevity, the coordinate label is omitted when the field is evaluated at the origin, $\psi\equiv \psi(0)$.
As usual, each bracket is a further order in the expansion. In the above we used the convention that terms with negative powers of the fields are actually absent, e.g. the term $|\psi|^{2(\ell-2)}$ must be dropped for $\ell=1$.
As for the effective action, the first-order corrections are strictly local, while the following orders carry nonlocal (short-ranged) corrections mediated by $F(x)$ and $R(x)$.
In Fig. \ref{fig_LL_onept} we compare the semiclassical expansion of the local observables with the exact results presented in Sec. \ref{sec_integrability}, finding excellent agreement.

\begin{figure}[t!]
\includegraphics[width=0.8\columnwidth]{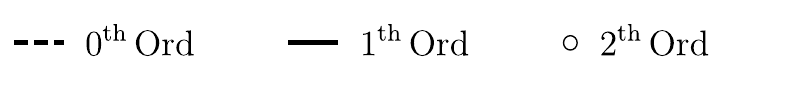}\\
\includegraphics[width=0.95\columnwidth]{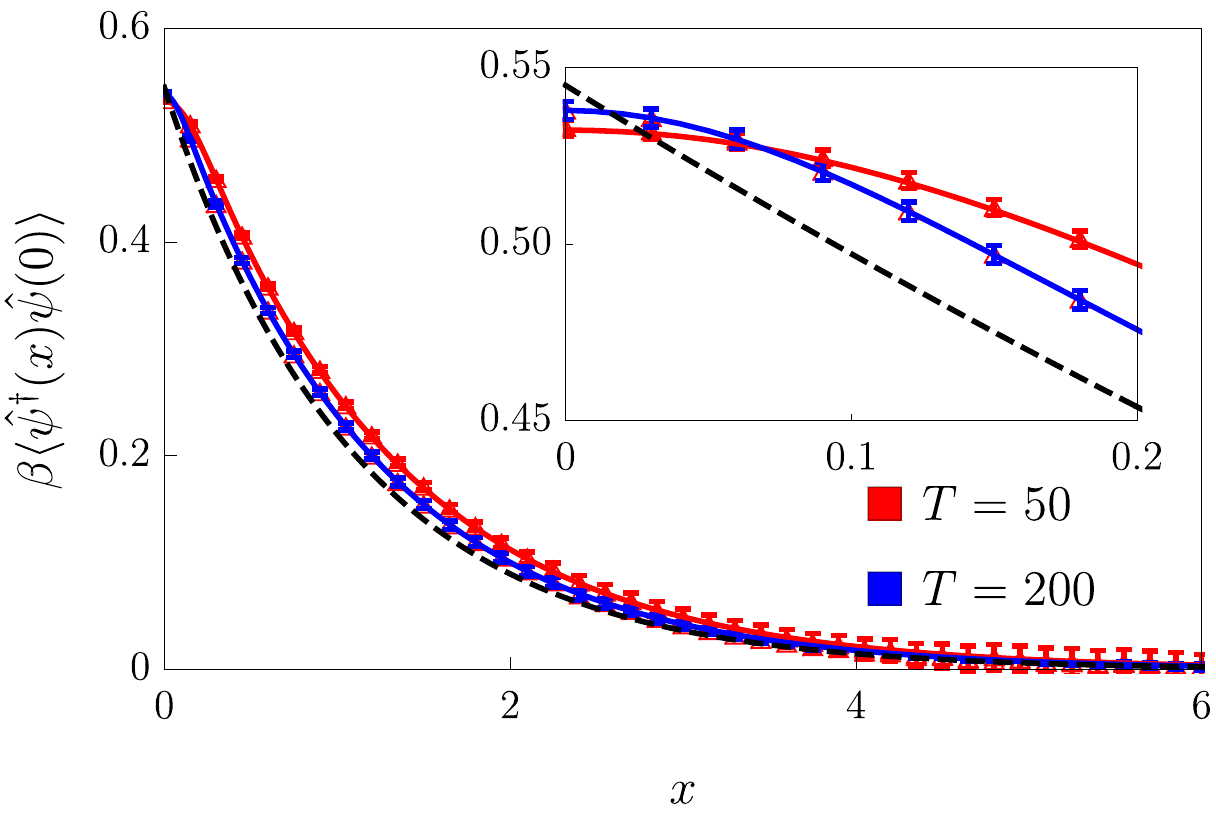}
\caption{\label{fig_1ptdm}
We consider the semiclassical expansion of the one-body density matrix $\beta\langle \hat{\psi}^\dagger(x)\hat{\psi}(0)\rangle$ (which is real and symmetric w.r.t. $x\to-x$) for two illustrative different temperatures $T=50$ and $T=200$. While appreciable corrections to the classical result (dashed line) are introduced including the first-order quantum corrections (solid lines), considering the second-order quantum corrections (symbols) does not change the result on an appreciable scale. This suggests that the first-order corrections are sufficient to approach the true quantum expectation values with good precision.
As discussed in the text, quantum corrections are essential to smoothen the nonanalyticity at the origin (inset).
}
\end{figure}

Having checked the one-point functions, we now revert to quantities where no integrability results are available. One prominent example it the one-particle density matrix, $\mathcal{O}^{\text{1dm}}[\sqrt{\beta}\hat{\psi}]=\beta\langle \hat{\psi}^\dagger(x)\hat{\psi}(y)\rangle$, its expansion up to the second order being
\begin{multline}\label{O_eff_oneppt}
\mathcal{O}_\text{eff}^{\text{1dm}}[\psi]=\Bigg[\psi^*(x)\psi(y)\Bigg]+
\Bigg[C(x-y)\Bigg]\\
+\Bigg[-4c_\text{cl} (|\psi|^2-d)\int \dd z\, F\left(\frac{|x-z|+|y-z|}{2}\right)\Bigg]+...
\end{multline}
As usual, square brackets are used to separate different orders of the expansion.
In Fig. \ref{fig_1ptdm}, we plot the semiclassical expansion results for $\langle \mathcal{O}^{\text{1dm}}[\sqrt{\beta}\hat{\psi}]\rangle$, since there are no analytical or numerical exact results to be used as benchmarks, comparing different orders of the expansion is of utmost importance for checking the reliability of the approximation. 

Quantum corrections are particularly important for the short-distance behavior
of the one-body density matrix. Indeed, the classical result is not differentiable at the origin. In other words, considering the momentum distribution
$n(k)=\int \dd x e^{-i kx }\langle \hat{\psi}^\dagger(x)\hat{\psi}(0)\rangle$,
in the classical limit one finds $n(k)\sim k^{-2}$ for large
$k$. These ``fat tails" of the momentum distribution eventually cause the UV
catastrophe, similarly to the famous example of the black body radiation: the
classical expectation value of the energy density is UV-divergent.  On the other hand,
the large $k$ behavior of the momentum density $n(k)$ in the quantum model is
known to decay as $n(k)\sim \mathcal{C}/k^{4}$
\ocite{PhysRevLett.91.090401,TAN20082952,PhysRevLett.110.020403,PhysRevLett.121.220402,PhysRevLett.121.220402}; furthermore the power law decay is linked to the average of the interaction term 
$\mathcal{C}=m
c^2\langle\hat{\psi}^\dagger\hat{\psi}^\dagger\hat{\psi}\hat{\psi} \rangle$.
Within our expansion, the first-order correction in Eq. \eqref{O_eff_oneppt}
is crucial in converting the classical result $n(k)\sim k^{-2}$ into the quantum
behavior. Indeed, in the Fourier space the function $C$ \eqref{eq_C_f}
exhibits the $\sim k^{-2}$ decay with the opposite sign when compared to the
classical ``fat tails"" arising from $\langle \psi^*(x)\psi(y)\rangle$, leading to cancellation. From the computational point of view, a short distance study of the one-body density matrix  is greatly affected by the finite lattice space in the numerical implementation. Therefore a careful extrapolation is needed for the continuum model. Such an extensive study would go beyond the purpose of the current work, hence we simply point out that inclusion of the quantum corrections smooths the one-body density matrix.

Finally, we apply the semiclassical expansion to study the FCS of the number of particles on an interval. 
So far some results for the FCS of the number operator in the LL model have been obtained only in the classical limit \ocite{PhysRevLett.122.120401} (see also Ref. \ocite{vecchio2020exact} for FCS of the density operator in the classical limit and out-of-equilibrium), in the quantum case for very small intervals \ocite{PhysRevLett.120.190601,Bastianello_2018}. A related problem of emptiness formation probability in the LL model was investigated using the Luttinger liquid techniques in \ocite{ABANOV2002565,alex2005hydrodynamics}. 
Let us consider the observable $\mathcal{N}^{(L)}[\sqrt{\beta}\hat{\psi}]$ defined as
\be\label{eq_NL}
\mathcal{N}^{(L)}[\sqrt{\beta}\hat{\psi}]=\frac{\beta}{\sqrt{L}}\int_0^L\dd x\, [\hat{\psi}^\dagger(x)\hat{\psi}(x)-\langle\hat{\psi}^\dagger(x)\hat{\psi}(x)\rangle]\, .
\ee

We define the FCS as $P_{\mathcal{N}^{(L)}}(w)=\langle \delta(w-\mathcal{N}^{(L)}[\sqrt{\beta}\hat{\psi}])\rangle$, and use the semiclassical expansion to access its generating function, $G_{\mathcal{N}^{(L)}}(\lambda)=\langle e^{i\lambda \mathcal{N}^{(L)}[\sqrt{\beta}\hat{\psi}]}\rangle$, i.e. the Fourier transform of the FCS. The particular normalization of $\mathcal{N}^{(L)}$ is chosen by considering the limit of large intervals. Indeed, for intervals much larger than the correlation length the central limit theorem is expected to hold with the FCS approaching the Gaussian distribution, $\lim_{L\to \infty} P_{\mathcal{N}^{(L)}}(w)=\frac{1}{\sqrt{2\pi}\sigma}e^{-\frac{w^2}{2\sigma^2}}$, with $\sigma$ depending on both the temperature and the chemical potential.
\begin{figure*}[t!]
\includegraphics[width=0.9\columnwidth]{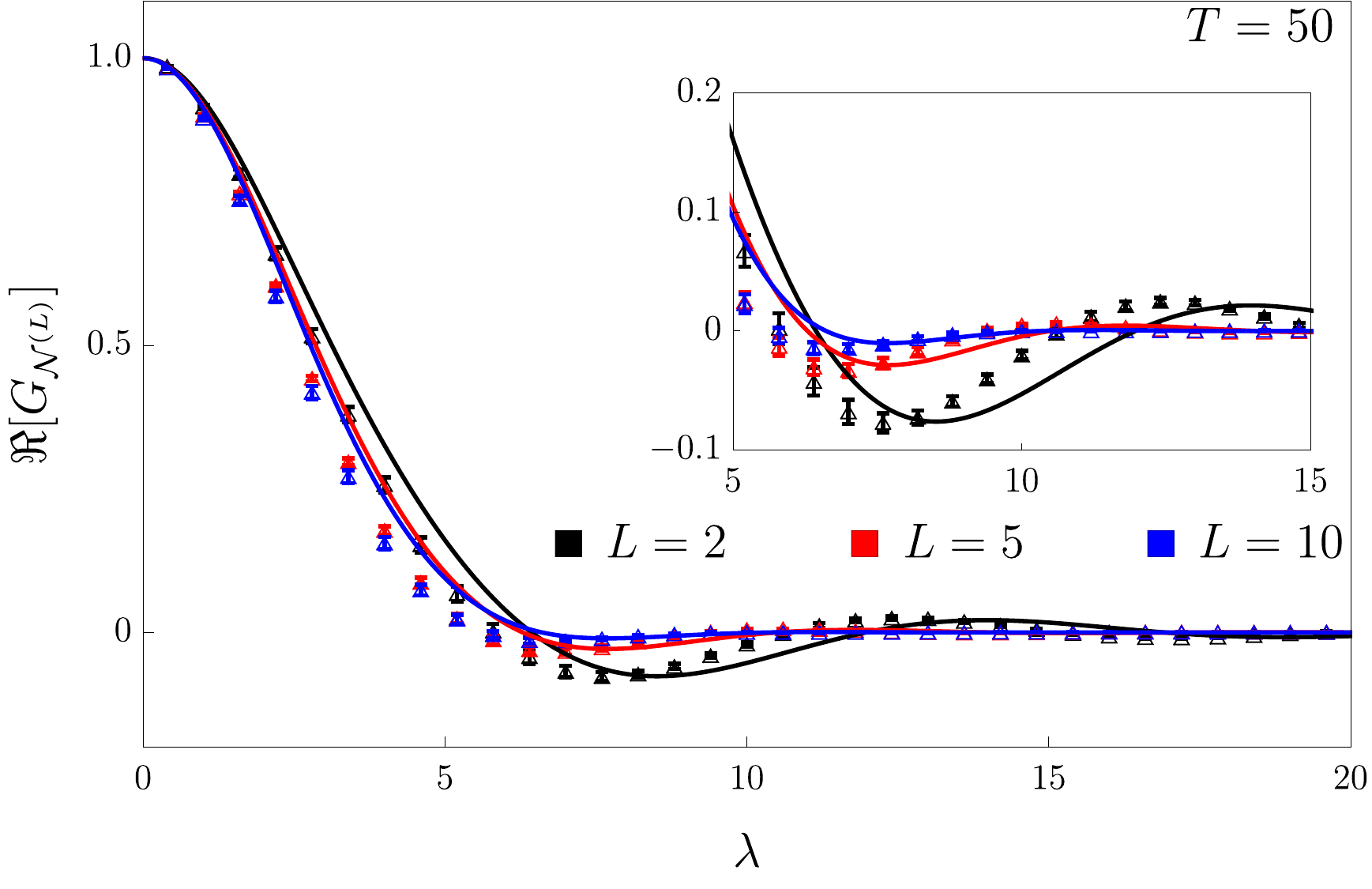}\hspace{2pc}\includegraphics[width=0.9\columnwidth]{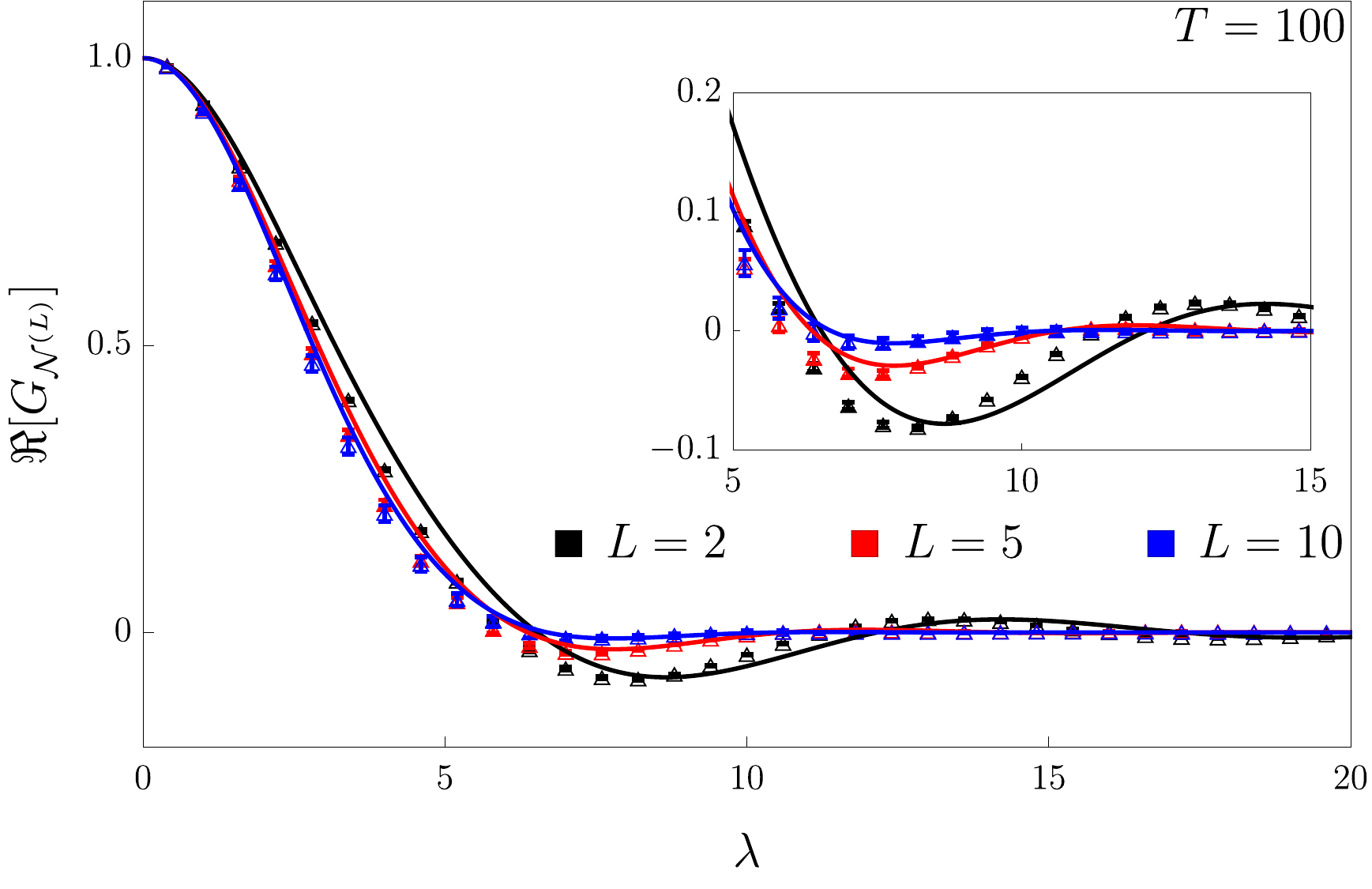}\\
\includegraphics[width=0.9\columnwidth]{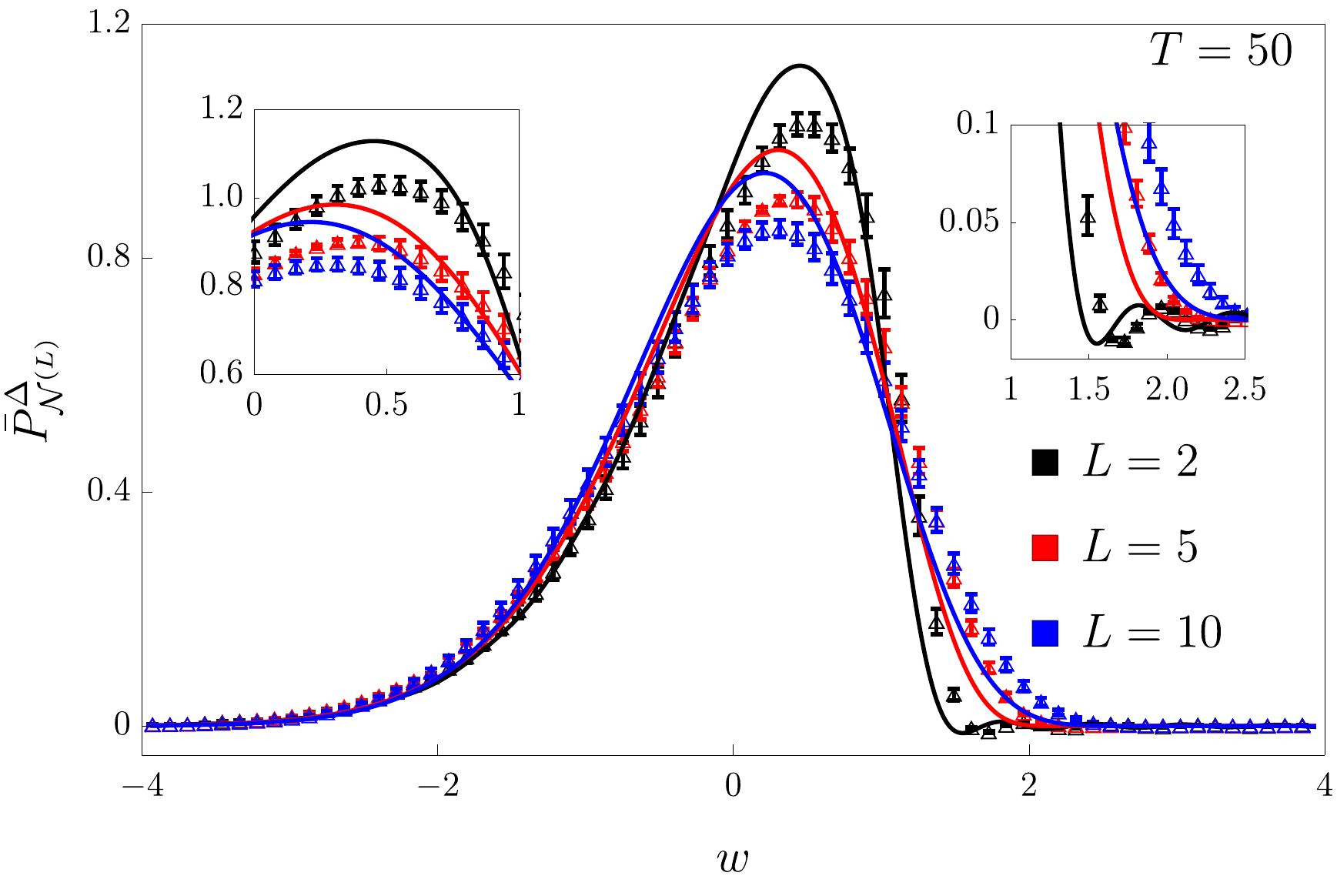}\hspace{2pc}\includegraphics[width=0.9\columnwidth]{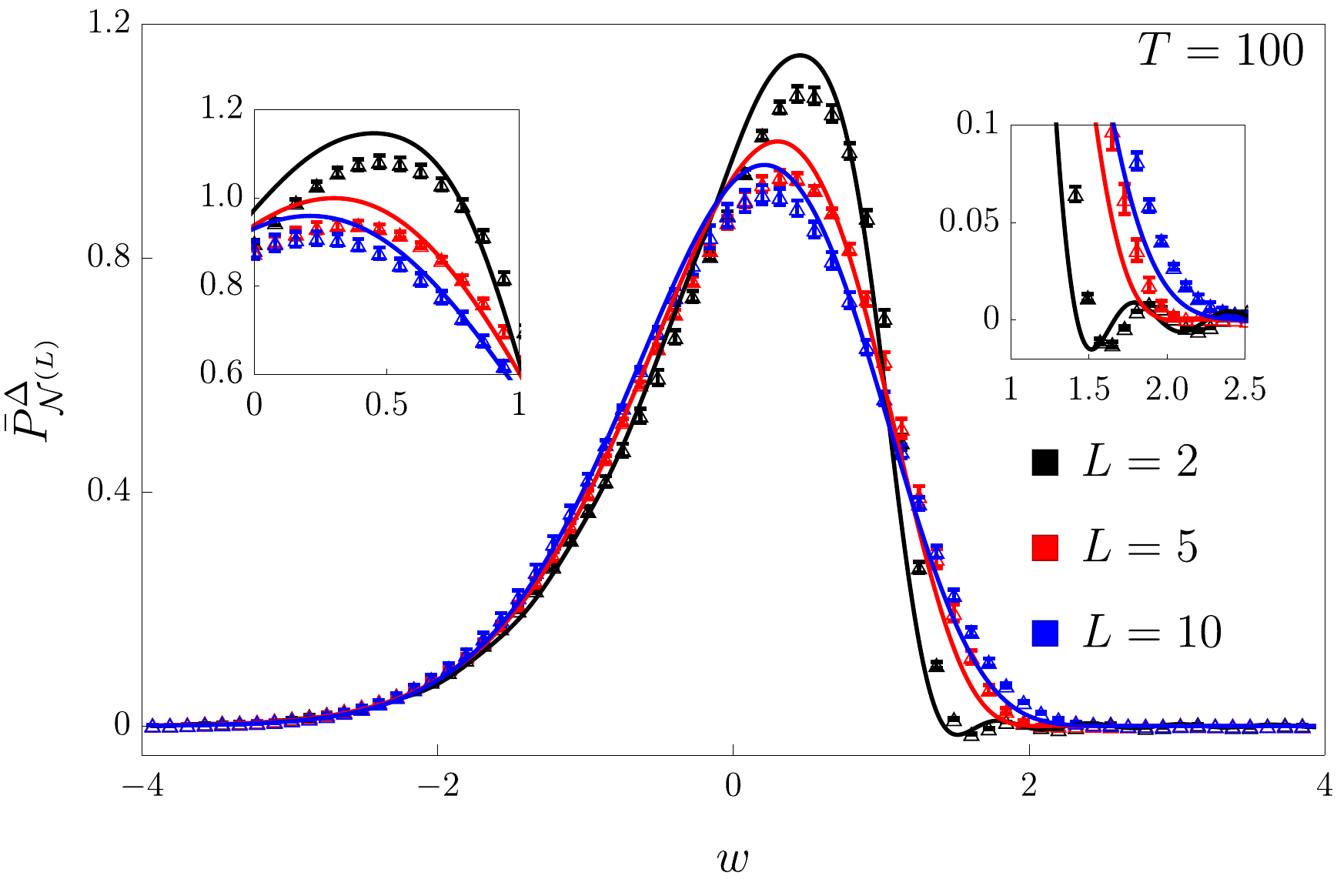}\\
\caption{\label{fig_fcs_LL}
We consider the generating functions and the FCS of the number of particles Eq. \eqref{eq_NL} for different interval sizes and temperatures, comparing the semiclassical approximation results (solid lines) to the first-order quantum corrections (symbols).
(Top) Real parts of the generating functions. (Insets) Zoom to appreciate the differences between the classical results and the first-order corrections.
(Bottom) FCS Eq. \eqref{eq_Pbar} smoothened according to Eq. \eqref{eq_cg_fcs} with $\Delta^{-1}=20$. (Insets) Zooms on particular regions. The finite cut-off causes the appearance of negative values due to the nonconstant sign of the kernel Eq. \eqref{eq_cg_fcs}. The presence of negative values of $\bar{P}^{\Delta}_{\mathcal{N}^{(L)}}$ is particularly evident in the case of the smallest interval $L=2$, where the hard cut-off on the generating function matters the most. 
}
\end{figure*}

Our  semiclassical expansion formalism utilizes the path integral representations of thermal states, therefore quantum expressions must be properly normal-ordered before being fed into the semiclassical formalism as observables.
In the case of the number operator normal-ordering simply gives \ocite{PhysRevB.95.174303,PhysRevA.89.013609,Mazza_2014,PhysRevA.97.033609}
\be
G_{\mathcal{N}^{(L)}}(\lambda)=e^{-i\lambda \beta\sqrt{L}\langle \hat{\psi}^\dagger\hat{\psi}\rangle}\langle :e^{(e^{i\lambda\beta/\sqrt{L}}-1)\int_0^L\dd x\, \hat{\psi}^\dagger(x)\hat{\psi}(x) }:\rangle\, .
\ee

Notice that the number of particles on an interval is an integer number, hence the operator $\mathcal{N}^{(L)}$ \eqref{eq_NL} has a discrete spectrum. This is reflected into a periodic structure of the generating function, $G_{\mathcal{N}^{(L)}}(\lambda+2\pi\sqrt{L}/\beta)=e^{-i2\pi  \langle \hat{\psi}^\dagger\hat{\psi}\rangle}G_{\mathcal{N}^{(L)}}(\lambda)$, which, after taking the Fourier transform, results in a Dirac comb in the FCS
\be\label{eq_fcs_diracComb}
P_{\mathcal{N}^{(L)}}(w)=2\pi \delta (e^{i2\pi  [ (\sqrt{L}/\beta)w +\langle\hat{\psi}^\dagger\hat{\psi}\rangle]}-1)\bar{P}_{\mathcal{N}^{(L)}}(w)\, ,
\ee 
with $\bar{P}_{\mathcal{N}^{(L)}}(w)$ being the smooth part
\be\label{eq_Pbar}
\bar{P}_{\mathcal{N}^{(L)}}(w)=\int_{-\pi\sqrt{L}\beta^{-1}}^{\pi\sqrt{L}\beta^{-1}}\frac{\dd \lambda}{2\pi}e^{-i\lambda w}G_{\mathcal{N}^{(L)}}(\lambda)\, .
\ee

Analogously to the anharmonic oscillator of Sec. \ref{sec_anh_Fey}, the generating function can be represented as
\be\label{eq_gen_fcs_LL}
G_{\mathcal{N}^{(L)}}(\lambda)=e^{-i\lambda \beta\sqrt{L}\langle \hat{\psi}^\dagger\hat{\psi}\rangle}\frac{\int \mathcal{D} \psi\, e^{-\mathcal{S}^{\mathcal{N}^{(L)},\lambda}_\text{eff}[\psi]}}{\int \mathcal{D}\psi\, e^{-\mathcal{S}_\text{eff}[\psi]}}\, ,
\ee
where $\mathcal{S}^{\mathcal{N}^{(L)},\lambda}_\text{eff}[\psi]$ is defined as
\be
\mathcal{S}_{\text{eff}}^{\mathcal{N}^{(L)},\lambda}[\psi]=-\sum_{j=1}^\infty\frac{1}{j!}\left\langle\left(\frac{e^{i\lambda\beta/\sqrt{L}}-1}{\beta} \mathcal{N}^{(L)}-\mathcal{S}_\text{int}\right)^j \right\rangle_\text{free}^\text{c}\, .
\ee
Notice that, in comparison with Eq. \eqref{eq_eff_Olambda}, we replaced $i\lambda\to \beta^{-1}(e^{i\lambda\beta/\sqrt{L}}-1)$ due to the normal ordering.
The effective action $\mathcal{S}_{\text{eff}}^{\mathcal{N}^{(L)},\lambda}[\psi]$ can be now systematically expanded in the quantum corrections.
For the sake of simplicity, we only consider the first order of the expansion
\begin{multline}\label{eq_genfcsLL}
\mathcal{S}_{\text{eff}}^{\mathcal{N}^{(L)},\lambda}[\psi]-\mathcal{S}_\text{eff}[\psi]=\Bigg[-\frac{e^{i\lambda\beta/\sqrt{L}}-1}{\beta}\mathcal{N}^{(L)}[\psi]\Bigg]\\
+\Bigg[-\frac{e^{i\lambda\beta/\sqrt{L}}-1}{\beta}L C(0)\\
-\left(\frac{e^{i\lambda\beta/\sqrt{L}}-1}{\beta}\right)^2\int_{0}^L \dd x\dd y\,   \Re\big[C(x-y)\psi^\dagger(x)\psi(y)\big]\Bigg]+...
\end{multline}
Square brackets are used to point out the zeroth and the first orders, respectively, and $\mathcal{S}_\text{eff}[\psi]$ Eq. \eqref{eq_eff_S_LL} and $\langle \hat{\psi}^\dagger\hat{\psi}\rangle$ appearing in Eq. \eqref{eq_gen_fcs_LL} must be expanded to the same order for consistency.
The expansion of the generating function above is not expected to be valid for arbitrary values of $\lambda$, but it has a natural cutoff $|\lambda|\le \Delta^{-1} \ll \beta^{-1}\sqrt{L}$. Notice that in the limit of small temperatures and large intervals the $\lambda-$domain where the semiclassical expansion is valid is expected to grow: since $G_{\mathcal{N}^{(L)}(\lambda)}$ is a fast decaying function (see Fig. \ref{fig_fcs_LL}), this ultimately implies that the role of the cutoff $\Delta$ becomes negligible.
As was discussed in the case of the anharmonic oscillator Eq. \eqref{eq_cg_fcs}, imposing a hard cutoff on the generating function is equivalent to coarse-graining the FCS on an interval $\sim \Delta$.
In Fig. \ref{fig_fcs_LL}, we present the FCS for different temperatures and sizes of the interval, comparing the classical result with the first-order quantum corrections. In particular, we focus on the real part of the generating function (first row) for $\lambda>0$ (the function is symmetric). Appreciable differences between the quantum corrections and the classical approximation are visible (see insets). 
Then, we consider the Fourier transform of the generating function and access the full counting statistics. We consider a hard cut-off in the Fourier space, $\Delta^{-1}=20$, computing the associated smoothed version of the FCS, according with Eq. \eqref{eq_cg_fcs}. As expected, increasing $L$ the FCS approaches a Gaussian.
Notice that the smoothed FCS $\bar{P}^{\Delta}_{\mathcal{N}^{(L)}}$, contrary to the $\bar{P}_{\mathcal{N}^{(L)}}$, is not guaranteed to be positive due to the presence of the oscillating kernel Eq. \eqref{eq_cg_fcs}. Indeed, we observe the presence of small negative values (see inset). Such oscillations observed in $\bar{P}^{\Delta}_{\mathcal{N}^{(L)}}$ are consequences of the hard cut-off rather than a true feature of the FCS.

Several comments are due when comparing the result of Ref. \ocite{PhysRevLett.122.120401} with the zeroth order of the semiclassical expansion here presented. Apart from the global rescaling and the shift in the definition of the number operator \eqref{eq_NL}, the classical field result of Ref. \ocite{PhysRevLett.122.120401} is obtained from Eq. \eqref{eq_genfcsLL} by keeping only the zeroth-order term of the expansion, and with the further approximation $\frac{e^{i\lambda\beta/\sqrt{L}}-1}{\beta}\simeq i\lambda/\sqrt{L}$, which is the dominant contribution in the high-temperature limit. Because of this further approximation the classical field result of Ref. \ocite{PhysRevLett.122.120401} does not exactly coincide with the zeroth order of Eq. \eqref{eq_genfcsLL}, but they do agree in the high-temperature limit.

\section{Conclusions}
\label{sec_conclusions}

We presented a semiclassical approach to one-dimensional quantum many-body systems in thermal equilibrium. The classical limit is achieved in the high-temperature and weak interactions regime, but at any finite temperature quantum mechanics affects the classical approximation. We show how quantum corrections can be systematically taken into account by means of a renormalization of the energy and observables of the classical model.
For the sake of concreteness we focused on the 1D Bose gas with contact interactions; we benchmarked our approach using exact results obtained from the integrability.

\begin{figure}[t!]
\includegraphics[width=1\columnwidth]{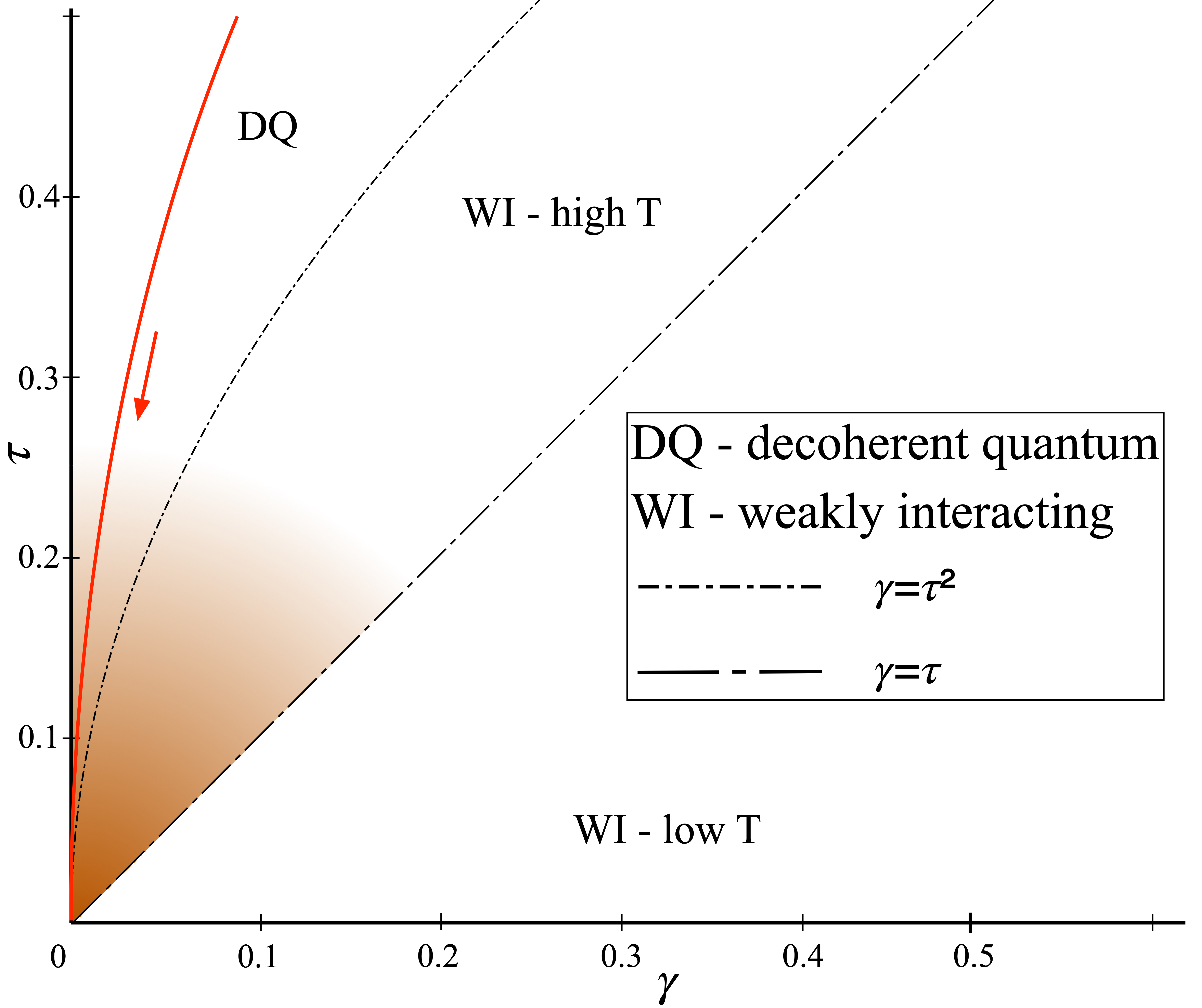}
\caption{\label{fig_validity}Regions of 1D Bose gas marked on the
  $\tau$-$\gamma$ plane, where $\tau$ is the ratio of $T$ and the temperature
  of quantum degeneracy, $\tau=T/T_d=m k_B T/\hbar^2 n^2$, and $\gamma=mc/
  \hbar^2 n$ is the 1D interaction parameter. The regimes are classified
  according to Ref. \ocite{PhysRevLett.100.160406}, with the relevant regimes being decoherent
  quantum (DQ), $\tau\ll1$, $\gamma \ll \tau^2$; weakly interacting (WI - hight $T$),
  $\gamma \ll \tau \ll \sqrt{\gamma}$; and weakly interacting (WI - low $T$), $\tau \ll
  \gamma \ll 1$. The classical field limit is attained at the origin,
  $\gamma,\tau\to 0$ by approaching it from either the high-$T$ WI or DQ regions.
  The semiclassical expansion is expected to be 
  valid for small but finite values of the parameters $\gamma$ and $\tau$, which is schematically represented by gradient color filling. As an example, the red solid line is shown, corresponding to the exact Lieb-Liniger solution for the $\mu=1$, $c_{cl}=1$, and $m=1/2$ case, for which the data is plotted on Fig. 4. The red arrow shows the direction of the classical field limit.}
\end{figure}

In particular, quantum corrections correctly capture the expectation values of the moments of the density operator, $\langle (\psi^\dagger)^n\psi^n\rangle$, in the high-temperature regime. We show how quantum corrections account for important deviations from the classical result with a rapid convergence to the exact value, which in turn is poorly approximated by the classical computation at any large, but finite, temperature.
We then applied our method to study experimentally relevant quantities which are not accessible by means of the state-of-the-art integrable techniques, namely the one-body density matrix $\langle \psi^\dagger(x)\psi(y)\rangle$ and the full counting statistics of the number of particles on an interval.
Our method is completely general and does not rely on the specific form of the interactions, hence it can be easily generalized to, e.g., finite-range interactions, or multicomponent gases.

The current study is immediately suitable for comparison with experimental and theoretical works studying 1D bosons with contact interactions from the ultracold atom perspective:
the connections of our parameters and those in the experiments can be easily obtained.
Going back to the dimensional units, the condition that $cT=c_\text{cl}$ and $\mu$ are kept fixed translates
into keeping the following dimensionless quantity constant:
\be
\left(\frac{\hbar^{2}}{2m}\left|\mu\right|\right)^{1/2}\frac{\left|\mu\right|}{k_{B}T}\frac{1}{c}.
\ee
In the classical limit, the quantity $(\hbar^{2}n^{2}\cdot\left|\mu\right|/2m)^{1/2}/k_{B}T$
is a constant as well, meaning that the average density $n$
scales linearly with temperature. Approaching the limit, the density
$n$ is still expected to scale almost linearly with $T$, the deviations
being determined by the quantum corrections. For example, the average
density for the range of temperatures considered in this work is given
by $(\hbar^{2}n^{2}\cdot\left|\mu\right|/2m)^{1/2}/k_{B}T\approx0.55$,
as can be read from Fig.~\ref{fig_LL_onept}.
Hence, one can easily  pinpoint  physical regimes of the gas as discussed in 
Ref.~\ocite{PhysRevLett.100.160406} by calculating  the 1D interaction parameter
$\gamma=mc/\hbar^{2}n\approx0.9\left(\left|\mu\right|/k_{B}T\right)^{2}$
and the dimensionless ratio
$\tau=mk_{B}T/\hbar^{2}n^{2}\approx1.7\left(\left|\mu\right|/k_{B}T\right)$
of  $T$  and the temperature of quantum degeneracy.
In Fig. \ref{fig_validity}, we schematically represent the validity region of the semiclassical expansion in the interacting Bose gas parameter space.

By further computing the coherence length of the gas, $l_{\phi}=\hbar^{2}n/mk_{B}T$, direct comparisons with Ref. \ocite{PhysRevLett.122.120401} can be made. For example, the case discussed on Fig. \ref{fig_fcs_LL} corresponds to the weakly interacting gas in the decoherent quantum regime, in the limit of large interval sizes $L/l_\phi\gg1$. 

Several interesting directions remain open for the future.
First of all, it would be interesting to compare our findings with actual experimental data: for instance, in Ref. \ocite{Schweigler2017} two tunneling-coupled Bose gas are effectively described by the sine-Gordon model, classical treatment of which proved to be in good agreement with experimental observations. Studying the effects of quantum corrections is extremely interesting.

A natural direction is to address out-of-equilibrium setups, for example including quantum corrections to the stochastic Gross-Pitaevskii approach \ocite{Gardiner_2002,Gardiner_2003,Cockburn2009}.
Another interesting challenge at the interface of out-of-equilibrium and integrability is determining the steady-state after a quantum quench \ocite{PhysRevLett.96.136801}. In principle, in integrable models, the steady-state is completely determined by the expectation value of the conserved charges (or their generating function) in the initial state. However, in most of cases this problem is extremely hard to solve \ocite{Calabrese_2016}: very recently an efficient numerical method has been devised for the classical limit, allowing to extract the steady state directly from the pre-quench state \ocite{vecchio2020exact}. It would be extremely interesting to put quantum corrections back in the game in the spirit of the present work.

\subsection*{Acknowledgements}
We are grateful to F. S. M{\o}ller for helpful discussions and pointing out relevant experimental references.
AB acknowledges support from the European Research Council (ERC) under ERC Advanced grant 743032 DYNAMINT. MA acknowledges support from the China Postdoctoral Science Foundation under the grant 210633, the National Natural Science Foundation of China (NSFC) under the grant 11950410491, and XiÕan Jiaotong University Fundamental Research Fund under the grant 1191329138.

\appendix

\section{The transfer matrix approach}
\label{sec_TM}
One-dimensional classical field theories with local actions can be equivalently
formulated in terms of single-particle quantum mechanics, this is
known as the transfer matrix (TM) method  \ocite{PhysRevB.6.3409,Krumhansl1975,doi:10.1080/09500340008232189}. By omitting the $n\neq0$ Matsubara modes in the expansion of $\psi_E(x)$, which is equivalent to neglecting the Euclidean time dependence of the fields in the quantum action Eq. \eqref{Euclidean_action}, directly yields the classical field action
\begin{equation}
\mathcal{S}_{\text{cl}}\left[\psi\right]=\int\mathrm{d}x\,\left\{\frac{\left|\partial_{x}\psi\right|^{2}}{2m}+c_{\text{cl}}\left|\psi\right|^{4}-\mu\left|\psi\right|^{2} \right\}\,,\label{eq:ClassicalFieldAction}
\end{equation}
while the quantum corrections are being ignored. Relabelling the dummy integration variable having the meaning of imaginary time $t\equiv x$, and making further suggestive identifications,
\begin{equation}
\psi=re^{i\phi},\qquad\psi^{*}=re^{-i\phi},\label{eq:ChangeVariables}
\end{equation}
where $r$ and $\phi$ are the polar coordinates of a two-dimensional particle.
makes it obvious that the classical field action \eqref{eq:ClassicalFieldAction} also describes a quantum particle
of mass $M=1/m$ moving in the 2D quartic potential,
\begin{multline}\label{eq:EffectiveQMAction}
\mathcal{S}_\text{cl}\left[\psi\right]\\
\equiv\int\mathrm{d}t\,\left\{\frac{M}{2}\left(\frac{\mathrm{d}r}{\mathrm{d}t}\right)^{2}+\frac{Mr^2}{2}\left(\frac{\mathrm{d}\phi}{\mathrm{d}t}\right)^{2}+
c_\text{cl}r^{4}-\mu r^{2} \right\}\, .
\end{multline}
For macroscopically large systems the interval of $t$ integration can be taken $(-\infty,\infty)$, corresponding to quantum statistical physics at zero temperature, which is entirely determined by the low-energy quantum states.

The Hamiltonian of the effective quantum problem is 
$$
\hat{H}=-\frac{1}{2M}\left( \frac{1}{r}\partial_{r}\left(r\partial_{r}\right)+\frac{1}{r^{2}}\partial^{2}_{\phi} \right)+V(r),
$$
\begin{equation}
V(r)=c_\text{cl}r^{4}-\mu r^{2}.\label{eq:PotentialEnergy}
\end{equation}
The corresponding 2D Schrodinger's equation is separable, and by introducing the
angular momentum quantum number $l=0,\pm1,...,\pm\infty$, such that
\begin{equation}
\left\langle r,\phi\right.\left|nl\right\rangle =\psi_{nl}(r,\phi)=R_{nl}\left(r\right)e^{il\phi},\label{eq:SeparableWF}
\end{equation}
results in a 1D eigenvalue problem
\begin{equation}
\left[ -\frac{1}{2M}\frac{1}{r}\partial_{r}\left(r\partial_{r}\right)+\frac{l^{2}}{2Mr^{2}}+V(r)\right] R_{nl}(r)=E_{nl}R_{nl}(r),\label{eq:1DEigenvalueProblem}
\end{equation}
where $n=0,1,2,\ldots$ enumerates the discrete spectrum of the bounded potential for the given
$l$. The essence of the TM method is finding the classical field-theoretic correlators by solving for the low energy eigenvalues and eigenstates of Eq. \eqref{eq:1DEigenvalueProblem}, numerically or otherwise. 

In particular, field-theoretic observables such as correlation
functions of the fields are found by expressing them in terms
of the eigenstates and eigenvalues of Eq. \eqref{eq:1DEigenvalueProblem}. For example, the partition function is expressed as
\begin{equation}
\mathcal{Z}=\int\mathcal{D}\psi \,e^{-\mathcal{S}_\text{cl}\left[\psi\right]}=\sum_{nl}e^{-E_{nl}D},\label{eq:PartitionFunction}
\end{equation}
where $D$ is the total length of the system, and the overall normalization factor has been omitted as irrelevant. For macroscopic systems,
$\mathcal{Z}\approx e^{-E_{00}D}$ with exponential accuracy, where
$E_{00}$ is the energy of the ground state of Eq. \eqref{eq:PotentialEnergy}.
Similarly, one-particle correlation function is given by 
\begin{equation}
\left\langle \psi^{*}(x)\psi(0)\right\rangle =\sum_{n}\left|\left\langle 00\right|r\left|n1\right\rangle \right|^{2}e^{-\left(E_{n1}-E_{00}\right)|x|},\label{eq:OneParticleCorrelator}
\end{equation}
again with exponential accuracy in $D$.

Moreover the classical field FCS can be accessed as well by evaluating its coarse-grained generating function \eqref{eq_gen_fcs_LL}. Within the classical field approximation, the effective action $\mathcal{S}_\text{eff}^{\mathcal{N}^{(L)}}$ \eqref{eq_genfcsLL} is a strictly local object, hence it is amenable to the transfer matrix treatment. As described in Ref. \ocite{PhysRevLett.122.120401}, the generating function of the FCS can be accessed by mean of the identity
\be
\frac{\int \mathcal{D} \psi\, e^{-\mathcal{S}^{\mathcal{N}^{(L)},\lambda}_\text{eff}[\psi]}}{\int \mathcal{D}\psi\, e^{-\mathcal{S}_\text{eff}[\psi]}}=\left\langle 00\right|e^{-\left(\hat{H}_{\lambda}-E_{00}\right)L}\left|00\right\rangle\, , 
\ee
which is then fed into Eq. \eqref{eq_gen_fcs_LL}. Above, the modified, non-Hermitian Hamiltonian $\hat{H}_{\lambda}$ is obtained from the transfer-matrix representation of the zeroth-order effective action Eq. \eqref{eq_genfcsLL}, namely
\be
\hat{H}_{\lambda}=-\frac{1}{2M}\left( \frac{1}{r}\partial_{r}\left(r\partial_{r}\right)+\frac{1}{r^{2}}\partial^{2}_{\phi} \right)+V_{\lambda}(r)\, .
\ee
with
\be
V_{\lambda}(\hat{r})=c_\text{cl}r^{4}+\left(\frac{e^{i\lambda\beta/\sqrt{L}}-1}{\beta\sqrt{L}}-\mu\right) r^{2}\, .\label{eq:PotentialEnergy}
\ee

In other words, the classical field probability generating function is calculated by evolving the original ground state of the equivalent quantum-mechanical problem with the modified Hamiltonian $\hat{H}_{\lambda}$ over ``time" $L$, and then calculating the overlap with the original ground state. This approach not only provides an alternative formulation for numerical computation of the generating function, but also serves as a convenient starting point for various approximations which result in analytic expressions for FCS in various regimes of temperature and interval size.

\section{The Metropolis-Hastings algorithm}
\label{sec_metr}

The Metropolis-Hastings algorithm \ocite{10.1093/biomet/57.1.97} allows for a systematic sampling of the phase-space density of the classical theory in equilibrium.
The idea is to construct a suitable ergodic random walk in the phase space, then the thermal expectation values of the observables can be replaced with the averages along the evolution.
The interested reader can refer to Ref. \ocite{doi:10.1080/00031305.1995.10476177} for a detailed discussion of the method, here we provide a short summary of the algorithm, then apply it to our case of interest.
Let us assume that we are interested in a system of $M$ complex variables $\{\psi_i\}_{i=1}^M$ (which later on will be a lattice discretization of the classical field), let us also assume the probability for a certain field configuration $p[\{\psi_i\}_{i=1}^M]$ is of the form
\be\label{eq_p_met}
p[\{\psi_i\}_{i=1}^M]=\frac{1}{\mathcal{Z}}e^{-\mathcal{S}_\text{eff}[\{\psi_i\}_{i=1}^M]}\, ,
\ee
where the partition function $\mathcal{Z}$ is needed for normalization reasons, but its actual value is not important. We will refer to $\mathcal{S}_\text{eff}$ as the Metropolis energy.
Then, we give a dynamics to the system $\{\psi_i\}_{i=1}^M\to \{\psi_i'\}_{i=1}^M$ through the following rules. (1) Randomly choose a lattice site $j$ with equal probability. (2) Update the field configuration modifying the field on the chosen site $\psi_j\to \psi_j +\delta \psi_j$, the shift in the field can be chosen as random complex Gaussian variable of zero mean and variance $\chi=\langle |\delta\psi_j|^2\rangle$. The variance $\chi$ is a free parameter to be adjusted, as we discuss below. (3) Then, the new field configuration is accepted or rejected with some probability. This is determined computing the energy shift $\delta E=\mathcal{S}_\text{eff}[\{\psi_i'\}_{i=1}^M]-\mathcal{S}_\text{eff}[\{\psi_i\}_{i=1}^M]$. If $\delta E<0$ the new configuration is accepted, otherwise it is randomly accepted with probability $e^{-\delta E}$.

The above steps together constitute the fundamental update of the Metropolis evolution, which is then repeated. The variance $\chi$ must be tuned in such a way that, on average, the acceptance rate is roughly $0.5$. Apart from exceptional cases, any initial field configuration will converge towards the desired ensemble that is then sampled averaging the desired observables along the Metropolis evolution.
The most time-consuming step of the algorithm is computing the energy difference $\delta E$: the more degrees of freedom are coupled to the updated field $\psi_j$, the more demanding the computation of $\delta E$ is. From this perspective, including nonlocal terms in $\mathcal{S}_\text{eff}$ slows down the Metropolis evolution as the range of the interaction is increased.

In our case of interest, we discretize the continuum theory on a lattice $\psi(x=ja)\to\psi_j$, with $a$ being the lattice spacing. Integrals are represented by discrete summations and derivatives are replaced with the first-order increments, $\partial_x\psi(x=ja)\to a^{-1}(\psi_{j+1}-\psi_j)$. The lattice spacing is chosen small enough to attain convergence (within the statistical fluctuations intrinsic to the method) and the system large enough in order to avoid finite-size effects. In our simulations we used $a=0.03$ and $M=2000$, since this choice guaranteed us convergence both to the continuum and thermodynamic limits.
The error bars are estimated with the variance obtained from four independent samplings for each set of data.

For what concerns the generating function of the FCS, it is straightforwardly computed as per Eq. \eqref{eq_gen_fcs_LL} and noticing
\be
\frac{\int \dd^M \psi_i\, e^{-\mathcal{S}^{\mathcal{N}^{(L)},\lambda}_\text{eff}[\{\psi_i\}_{i=1}^M]}}{\int \dd^M\psi_i\, e^{-\mathcal{S}_\text{eff}[\{\psi_i\}_{i=1}^M]}}=\langle e^{\mathcal{S}_\text{eff}[\{\psi_i\}_{i=1}^M]-\mathcal{S}^{\mathcal{N}^{(L)},\lambda}_\text{eff}[\{\psi_i\}_{i=1}^M]} \rangle_p\, ,
\ee
where the l.h.s. is the discretized version of the analog expression in Eq. \eqref{eq_gen_fcs_LL}, on the r.h.s. with $\langle ...\rangle_p$ we denote the average with respect to the probability distribution Eq. \eqref{eq_p_met}. Then, the function $ e^{\mathcal{S}_\text{eff}[\{\psi_i\}_{i=1}^M]-\mathcal{S}^{\mathcal{N}^{(L)},\lambda}_\text{eff}[\{\psi_i\}_{i=1}^M]}$ is regarded as a $\lambda$ and $L$ dependent observable and its value is sampled along the Metropolis evolution.

\bibliography{biblio}

\end{document}